# When did Life Likely Emerge on Earth in an RNA-First Process?

Steven A. Benner,*[a,e] Elizabeth A. Bell,[b] Elisa Biondi,[a] Ramon Brasser,[c] Thomas Carell,[d] Hyo-Joong Kim,[e] Stephen J. Mojzsis,[f] Arthur Omran,[g] Matthew A. Pasek,[g] and Dustin Trail[h]

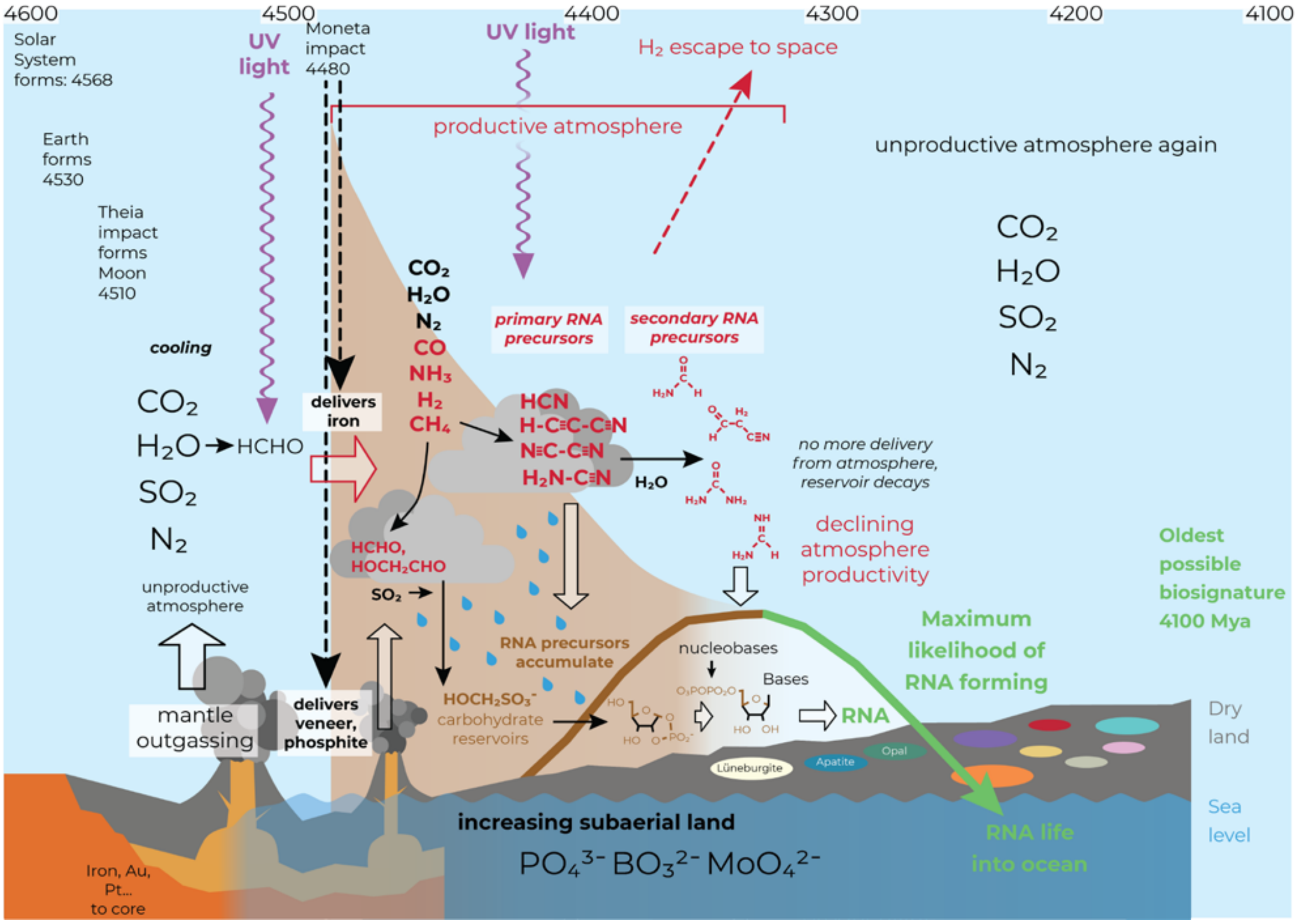

[a] Foundation for Applied Molecular Evolution, Alachua FL, USA, ebiondi@ffame.org; sbenner@ffame.org
[b] Department of Earth, Planetary, and Space Sciences, University of California, Los Angeles, ebell21@ucla.edu
[c] Earth Life Science Institute, Tokyo Institute of Technology, Tokyo Japan, brasser_astro@yahoo.com
[d] Fakultät für Chemie und Pharmazie, Ludwig-Maximilians-Universität, München, Deutschland thomas.carell@cup.lmu.de
[e] Firebird Biomolecular Sciences LLC, Alachua FL, USA, hkim@firebirdbio.com
[f] Department of Geological Sciences, University of Colorado, Boulder CO USA, Stephen.Mojzsis@colorado.edu
[g] School of Geosciences, University of South Florida, Tampa, FL, USA., arthur.omran@hotmail.com, mpasek@usf.edu
[h] Department of Earth and Environmental Sciences, University of Rochester, Rochester NY USA dustin.trail@rochester.edu
* manuscripts@ffame.org

**Abstract:** The widespread presence of ribonucleic acid (RNA) catalysts and cofactors in Earth's biosphere today suggests that RNA was the first biopolymer to support Darwinian evolution. However, most "path-hypotheses" to generate building blocks for RNA require reduced nitrogen-containing compounds not made in useful amounts in the $CO_2$-$N_2$-$H_2O$ atmospheres of the Hadean. We review models for Earth's impact history that invoke a single ~$10^{23}$ kg impactor (Moneta) to account for measured amounts of platinum, gold, and other siderophilic ("iron-loving") elements on the Earth and Moon. If it were the last sterilizing impactor, Moneta would have reduced the atmosphere but not its mantle, opening a "window of opportunity" for RNA synthesis, a period when RNA precursors rained from the atmosphere to land holding oxidized minerals that stabilize advanced RNA precursors and RNA. Surprisingly, this combination of physics, geology, and chemistry suggests a time when RNA formation was most probable, ~120 ± 100 million years after Moneta's impact, or ~4.36 ± 0.1 billion years ago. Uncertainties in this time are driven by uncertainties in rates of productive atmosphere loss and amounts of sub-aerial land.

## Introduction

The widespread presence of ribonucleic acid (RNA) cofactors and catalysts in today's terran biosphere supports the view that an early episode of life on Earth used RNA for both genetics and catalysis in an "RNA World".[1] This, in turn, led to the "Strong RNA First" hypothesis for the origin of Darwinian evolution on Earth.[2] Darwinian evolution involves a molecular/genetic line of descent with errors that do not anticipate future fitness, errors that are (i) removed by purifying selection if they are disadvantageous, (ii) selected naturally if they are adaptive, and (iii) fixed randomly if they have no impact on fitness. Darwinian evolution is thought to be necessary for non-intelligent matter to self-organize to produce behaviors that we ascribe to biology.

This strong hypothesis requires processes for the abiological formation of RNA.[3] Thus, many "path-hypotheses" have been offered to generate prebiotically the nucleos(t)ide building blocks needed for the assembly of RNA.[4] These are recently reviewed in Kitadai and Maruyama (2018).[5][a]

Others consider RNA formation to be difficult, and/or see the RNA-First model as simplistic. These scientists often suggest that Darwinian evolution began in non-RNA molecular genetic systems that were later replaced by RNA in a "grandfather's axe" scenario.[6][b] Still others suggest that Darwinian evolution was first achieved on Earth without a genetic biopolymer at all, under what is often called a "metabolism first" hypothesis.[8]

Under the RNA-First hypothesis, these path-hypotheses differ primarily in the details by which they create various bonds in RNA nucleos(t)ides, the number of discontinuities in their workflow, and the extent to which they follow a direct construction of the RNA molecule. For example, **Fig. 1** shows, without advocacy, a path-hypothesis that is relatively direct. It involves only precursors that might have been generated on Earth, exploits carbohydrate reservoirs stabilized by oxidized volcanic sulfur dioxide ($SO_2$),[9] uses amidophosphates as phosphorylating agents on arid land,[10] forms glycosidic bonds directly from ribose and nucleobases,[11] uses borate[12] to guide the regiochemistry of nucleoside phosphorylation,[4b] and obtains oxidized phosphate from borophosphate minerals that perform regiospecific phosphorlation.[13]

**Fig. 2**, again without advocacy, shows two path-hypotheses with indirect bond formation sequences. Here, bonds destined to join ribose to the pyrmidine and purine nucleobases are formed prior to complete assembly of the base, and embody a different strategy for creating phosphorylated species. **Fig. 3** shows, again without advocacy, a path-hypothesis that incorporates both oxidized and reduced nitrogen-containing species in the formation of RNA nucleobases. Another scheme, which focuses on reductive pathways to make higher carbohydrates relevant to **Fig. 2**, is shown in **Fig. 4**.

These figures represent, necessarily incompletely, a range of path-hypotheses that deliver RNA building blocks from primary precursors. However, their different details should not obscure their commonalities. All of these path-hypotheses involve relatively reduced organic molecules that serve as the precursors of the four standard RNA nucleobases (guanine, adenine, cytosine, and uracil) or "grandfather's axe" heterocycles (not shown).[7] Thus, all assume the production of substantial amounts of reduced primary precursors, likely in the Hadean atmosphere (before 4 billion years ago). These primary precursors include hydrogen cyanide (HCN), cyanamide ($H_2NCN$), cyanoacetylene (HCCCN), cyanogen (NCCN), ammonia ($NH_3$), and cyanic acid (HCNO). Further, they all assume that these (or their downstream products) avoided dilution into a global ocean, perhaps by adsorbing on solids,[14] or by delivery to sub-aerial land with a constrained aquifer.[15]

Historically over the past half-century, these reduced primary precursors have been considered "plausible" starting points for prebiotic chemistry.[16] All are seen by spectroscopy in star-forming nebulae.

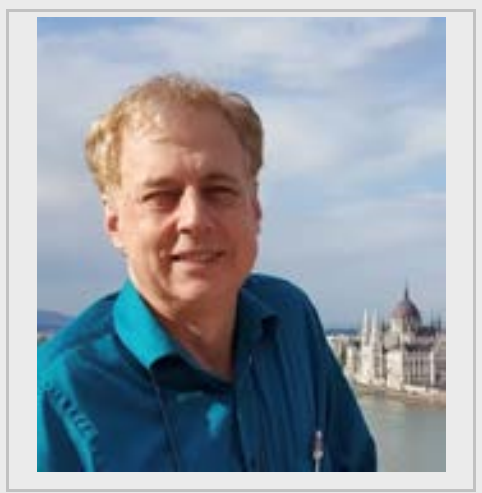

**Steven Benner** is a Distinguished Fellow at the Foundation for Applied Molecular Evolution, which he founded after serving at Harvard, the ETH Zurich, and the University of Florida. His research lies at the interface between natural history and the physical sciences. He has contributed to recent advances in synthetic biology and artificial life forms, paleogenetics resurrections of ancient genes and proteins, diagnostic and therapeutic medicine, chemical-mineral models for the origin of life, and planetary exploration in search of life forms.

[a] We omit many chemistry-related hypotheses that do not define specific paths that generate RNA building blocks. This omission should not be seen as a comment on the merits of those hypotheses.

[b] The "grandfather's axe" model holds that the dominance of RNA in the ribosome, RNA cofactors, RNAse P, and other processes inferred in the last universal ancestor of extant terran life is incorrectly used to infer that RNA supported the first Darwinian system. This model[7] holds that darwinian processes began with a different molecular system, whose components were then stepwise replaced by RNA, much as an old axe might have its head replaced, and then its handle replaced, all while continuing to function to chop wood. This alternative model, at least in its current form, does not prejudice the discussion here, since it also requires precursors that are also reduced organic compounds that need a reducing atmosphere to form.

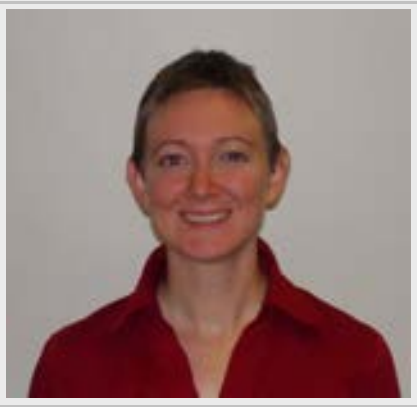

**Elizabeth Bell** studied geology at the University of South Carolina. After receiving her Ph.D. in geochemistry at the University of California, Los Angeles (UCLA), she received a Simons Collaboration on the Origins of Life postdoctoral fellowship. She is now a research scientist in the ion microprobe laboratory at UCLA. Her research concerns the earliest mineral records of Earth's crust and the emergence of life.

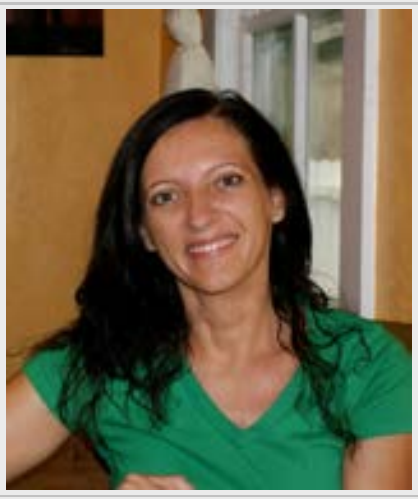

**Elisa Biondi** obtained her BS from the University of Bologna, working on Helicobacter methyltransferase systems. After receiving her PhD in Genetics and Astrobiology from the University of Florence, she moved to MU, developing HIV-aptamers for biotechnology applications and ribozymes relevant to the origins of life. She joined FfAME in 2012 and currently holds a double appointment as a senior research scientist at FfAME and Firebird. Her research interests include in-vitro evolution of nucleic acid libraries to deliver aptamers and ribozymes for astrobiology.

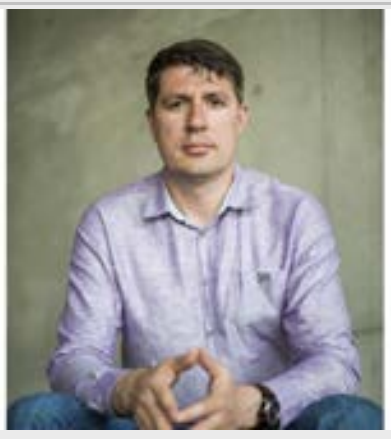

**Ramon Brasser** works as a planetary scientist at the Earth Life Science Institute in Tokyo, Japan. His primary research interest is to understand how rocky planets form and evolve, and which of these could develop a biosphere. To do so he makes use of state of the art numerical models and fuses these results with the latest cosmochemical data to place stringent constraints on how the planets formed and which of these could potentially develop a biosphere.

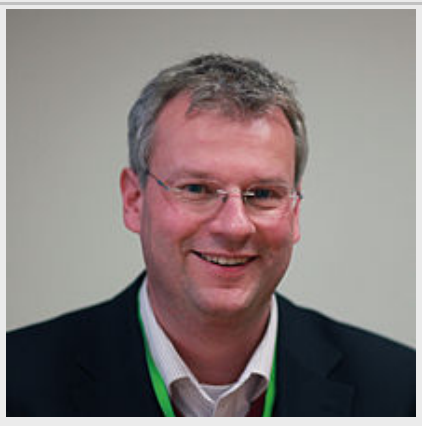

**Thomas Carell** completed his studies of chemistry in 1990 at the University of Münster, with a diploma thesis at the Max Planck Institute for Medical Research in Heidelberg. After his PhD work on porphyrin chemistry at that institute, he did postdoctoral at MIT. He finished his habilitation on DNA repair proteins at the ETH Zürich in 1998. From 2000 till 2004 he was Professor for Organic Chemistry at the University of Marburg, becoming then Professor for Organic Chemistry at the Ludwig Maximilians University of Munich. His main interest is DNA repair systems and chemistry related to the origins of life.

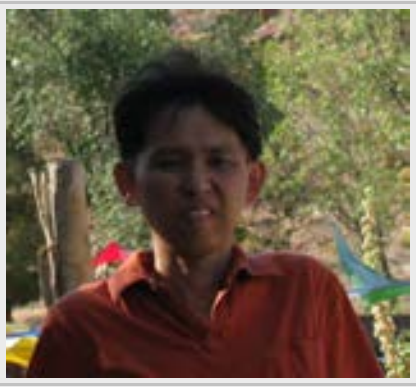

**Hyo-Joong Kim** received his B.S. and M.S. from the Seoul National University in South Korea. After spending 6 years in industry at LG Chem Ltd., he obtained a Ph.D. in chemistry from the University of Alabama (Tuscaloosa). He joined the FfAME/Firebird Biomolecular Sciences, LLC in 2006. His research activities include origin of life, especially abiologic synthesis of RNA building blocks, and synthetic biology based on artificial genetic materials.

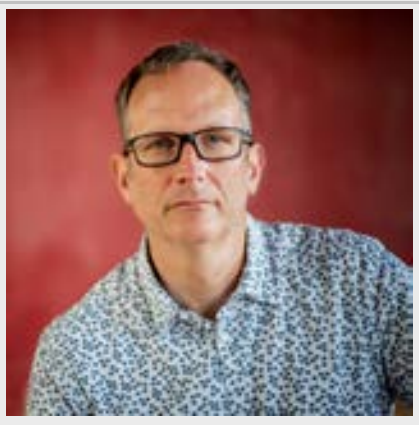

**Stephen J. Mojzsis** explores the fundamental physical and chemical aspects that govern the potential for a planet to harbor life, including field studies of early Archean terranes in Canada, Greenland, and Australia. His studies in early crustal process use geochronology and crystal chemistry with other isotopic techniques applied to minerals and rocks from Earth and the Moon, as well as Martian and asteroidal meteorites. Some of recent work investigates pre- and post-closure temperature thermal chemical diffusion modeling of accessory minerals in geochronology the evolution of atmospheric oxygen as it relates to evolution of biogeochemical cycles, and the overall physico-chemical considerations of the conditions for the origin of life on Earth and Earth-like planets.

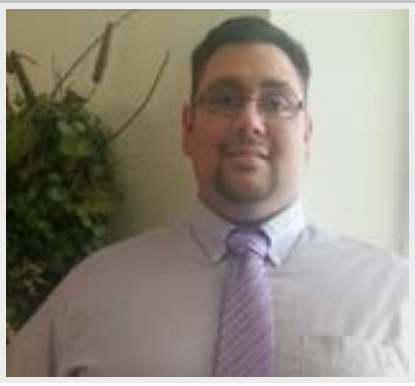

**Arthur Omran** studied at the University of North Florida, earning a BS in biology with a minor in World Religions and a MS in microbiology. He then studied at the Florida State University, earning a MS in biochemistry and a Ph.D. in physical chemistry. He is currently doing post-doctoral work with the Center for Chemical Evolution with Dr. Matthew Pasek at the University of South Florida. His research topics include the formose reaction, phosphorous availability, and the chemical evolution of RNA.

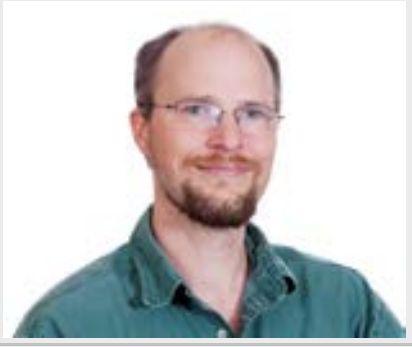

**Matthew Pasek** is a cosmochemist at the University of South Florida who researches the origin of life from the perspective of phosphorus chemistry. His research concerns redox changes to phosphorus that result from gas-solid reactions in the solar system to geochemical processing to biochemical modifications. His PhD is from the University of Arizona (2006, Planetary Science) and BS is from William and Mary (2002, Geology & Chemistry).

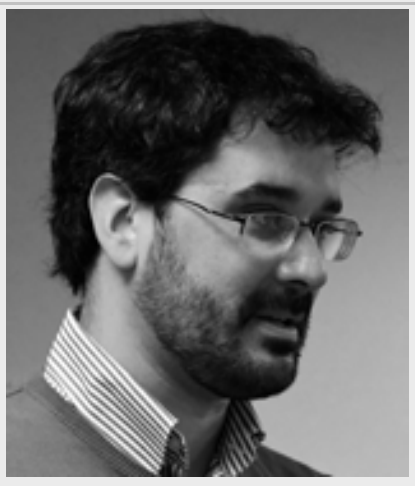

**Dustin Trail** is an Associate Professor in the Department of Earth and Environment Sciences at the University of Rochester. He studies the oldest known material from the formative stage of Earth, with ages that approach 4.4 billion years. He also designs and executes laboratory experiments to simulate planetary conditions of the nascent Earth. His contributions have led to new discoveries about the composition of the primordial crust, the atmosphere, and the nature of water-rock chemical reactions that occurred on Earth before 4 billion years ago.

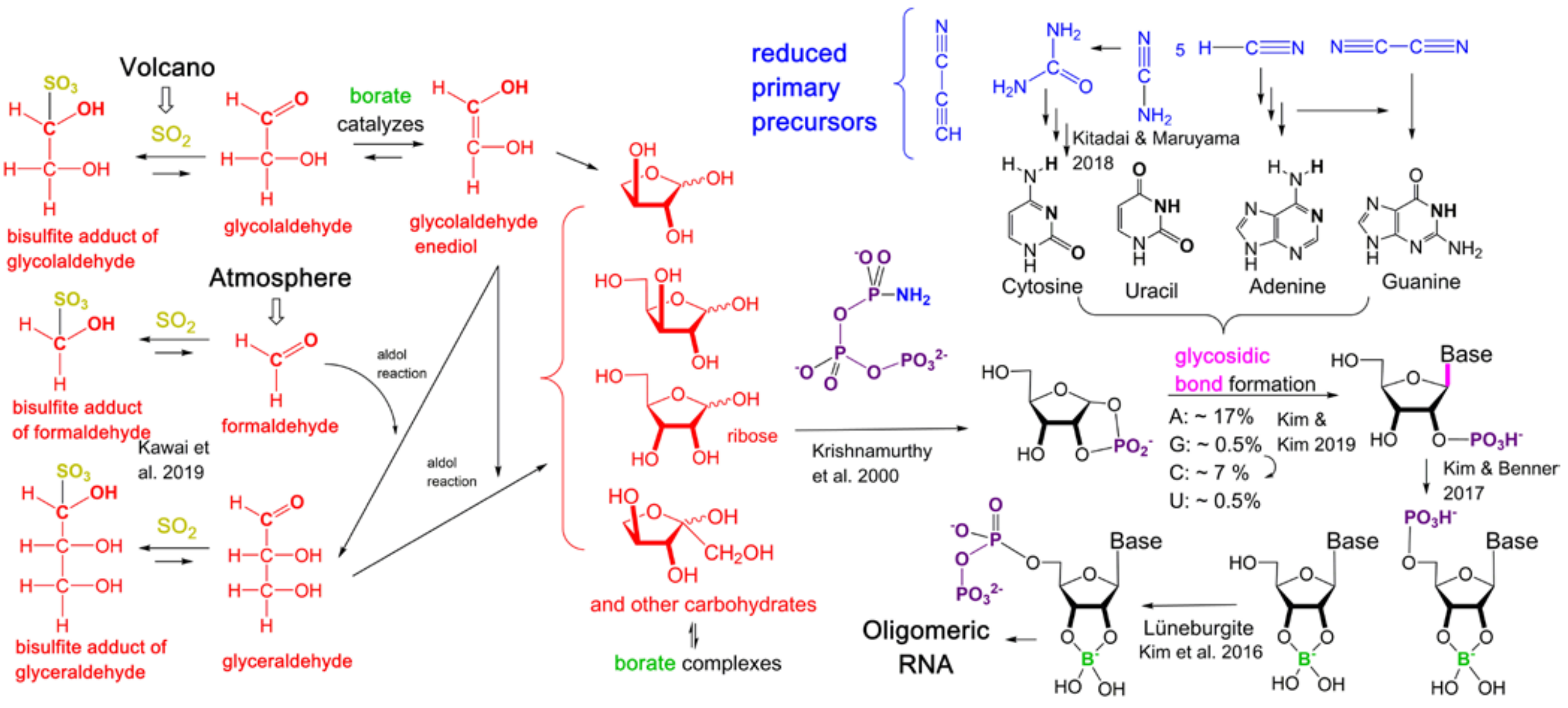

**Figure 1.** Schematic for a path-hypothesis that yields RNA nucleotides by direct joining of preformed canonical nucleobases to preformed ribose derivatives via a glycosidic bond (magenta).[11] The stereochemistry of various chiral molecules is arbitrary. This path-hypothesis invokes reservoirs of carbohydrates (red) arising from formaldehyde (HCHO) and traces of glycolaldehyde ($HOCH_2$-CHO) stabilized by $SO_2$ (yellow)[9] emerging from a Hadean mantle having an oxygen fugacity (redox state, $f_{O2}$) near the fayalite-magnetite-quartz buffer ($f_{O2}$ = FMQ -0.5 ± 2.3).[17] Black lines (left) indicate aldol reactions that carbohydrates undergo if released from their bisulfite adducts in the presence of borate and trimetaphosphate transformed with ammonia.[10] This path-hypothesis requires sub-aerial surfaces only intermittently submerged by water, or mechanisms (as yet unknown) that would concentrate the needed species from a global ocean on the Hadean Earth. Last, it requires HCN, HCCCN, $H_2NCN$, and other reduced atmosphere-generated primary precursors (in blue, not all shown). Formation of these depends strongly on the redox state of the atmosphere. For one critique of this path-hypothesis, see the Supplementary Information from Ritson et al. (2018).[18]

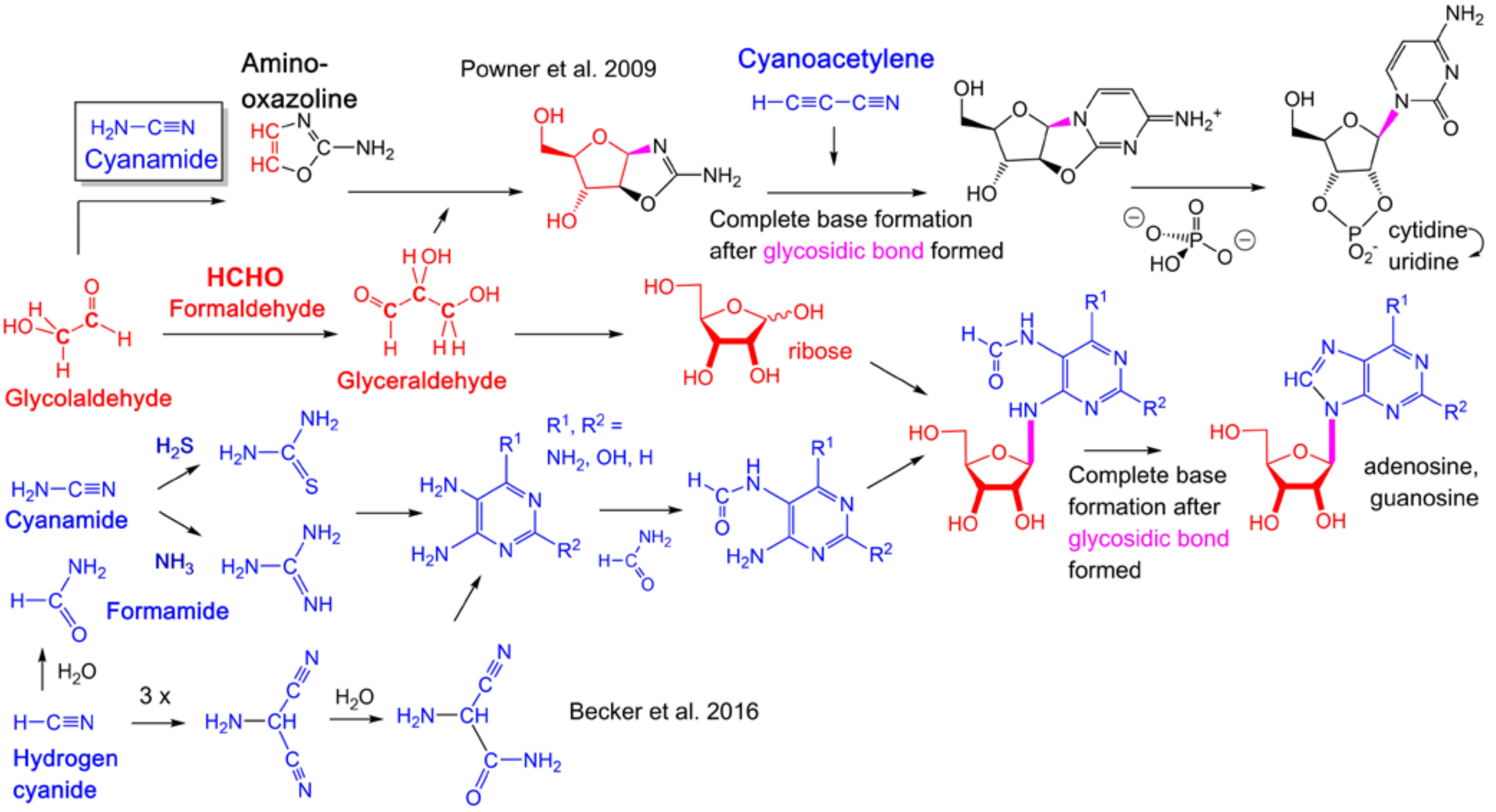

**Figure 2.** Schematics for two representative path-hypotheses forming RNA nucleos(t)ides, but where the bond destined to become the glycosidic bond (magenta) is formed *before* all of the atoms in the nucleobases are assembled. Adapted from Powner et al. (2009)[4a] and Becker et al. (2016).[4d] Stereochemistry is again entirely arbitrary. Carbohydrate precursors at the oxidation state of HCHO are in red. As in **Fig. 1**, HCN, HCCCN, $H_2NCN$, $H_2S$, $NH_3$, and other reduced primary precursors (blue) are needed as precursors for the heterocycles. Their formation again depends strongly on the redox state of the atmosphere.

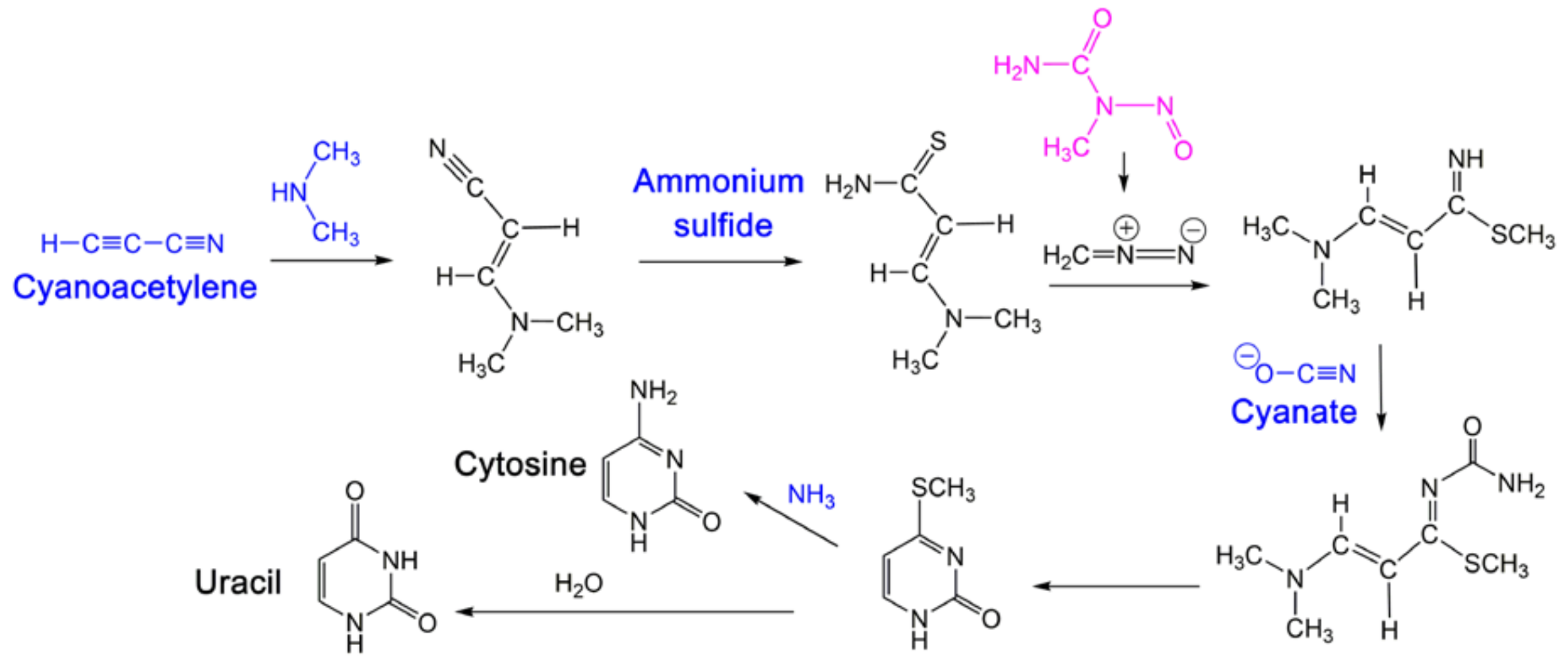

**Figure 3.** Schematic showing a path-hypothesis for the formation of advanced building blocks for RNA (here, cytosine and uracil) that involve both oxidized (magenta, note the N-O bond) and reduced (blue) nitrogen species. Other path-hypotheses, not shown, invoke reduced-oxidized mixtures, for example, to support the nitrosation of malonitrile (not shown here).[19] For models for the formation/destruction of various oxidized species in Hadean atmospheres, see Ranjan et al. (2019).[20] Adapted from Okamura et al. (2019).[19]

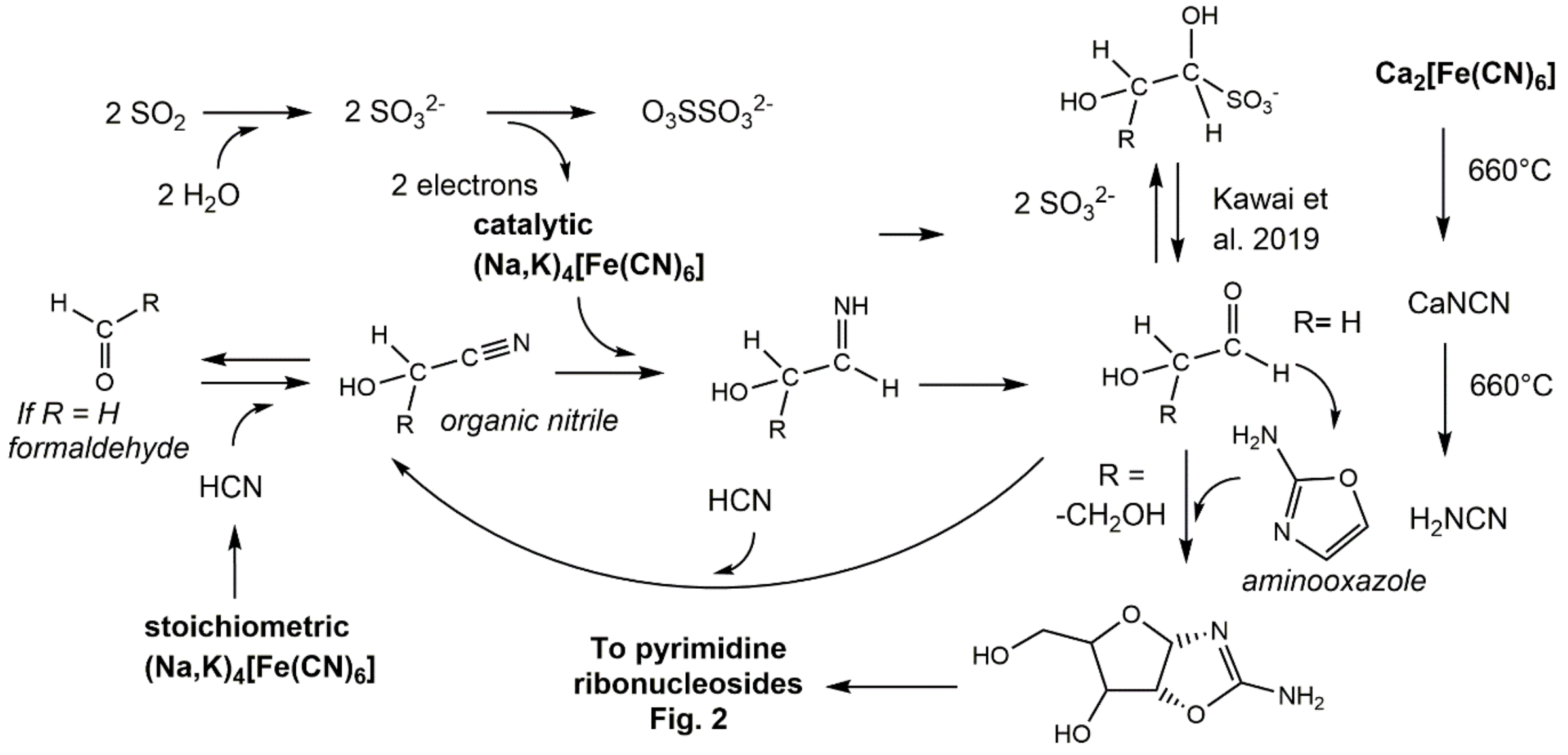

**Figure 4.** The "cyanosulfidic hypothesis" exploits the ability of ferrous iron to sequester cyanide, sulfite from volcanic $SO_2$ to serve as a reductant, and high energy photons, to allow sequential homologation of short linear carbohydrates to longer carbohydrates in a prebiotic analog of the Wolff-Kishner carbohydrate synthesiss. Reduced HCN, also proposed to be of atmospheric origin (here, from the late heavy bombardment) is used with metals ($Na^+$, $K^+$, $Ca^{2+}$) to create ferrocyanide,[21] which is proposed to serve as a source for HCN, a catalysts for the reduction of organic nitriles, and as a source of the cyanamide needed to generate an aminooxazole. Purine ribonucleosides would presumably be made by other processes. Adapted from Xu et al. (2018)[22] and Ritson et al. (2018).[18]

Unfortunately, this original argument for plausibility[23] lacks a good mechanism for these compounds to survive the harsh conditions associated with the planetary accretion processes that created Earth, in particular, the arrival of Theia, a ~$10^{25}$ kg body whose impact created our Moon 4.51 billion years ago (Ga). However, all can be generated by ultraviolet light, lightning, or silent electrical discharge acting on mixtures of gases in an atmosphere, if that mixture is sufficiently reducing.[5]

Many path-hypotheses also invoke secondary RNA precursors that arise via reaction of these reduced primary precursors with water.[24] These secondary precursors include formamide ($HCONH_2$) and ammonium formate ($NH_4^+$ $HCOO^-$) by hydrolysis of HCN; urea ($H_2NCONH_2$) by hydrolysis of $H_2NCN$; ammonia ($NH_3$) by hydrolysis of formamide, urea, and other species; cyanoacetaldehyde (NC-$CH_2$-CHO) by hydrolysis of HCCCN; and formamidine (HC(NH)$NH_2$) by reaction of $NH_3$ with $H_2NCN$; *inter alia*.

Some roles proposed for these primary and secondary precursors participating in the formation of RNA building blocks are listed in **Table 1**. These include the formation of nucleobases, carbohydrates, and other products having structures similar or identical to those of fragments of RNA.

Many path-hypotheses also posit roles for minerals[25] in advancing abiological precursors towards biological molecules. These include aluminosilicate clays[26] and micas.[27] **Table 2** collects some minerals that have been suggested to interact with various intermediates and products in various of the path-hypotheses in **Figs. 1-4**. Some are proposed as stabilizers, others are proposed as catalysts, while others are proposed as reagents.

**Table 1**. Reduced organic RNA precursors common to most prebiotic RNA synthesis models

| Formula | Compound | Role in RNA formation | Other roles |
|---|---|---|---|
| HCN | Hydrogen cyanide | Adenine precursor[28] | Source of $HCONH_2$, formate, $NH_3$ |
| $H_2NCN$ | Cyanamide | Dehydrating agent[29] | Source of $H_2NCONH_2$ by hydrolysis |
| HCCCN | Cyanoacetylene | Cytosine, uracil precursor[30] | Via cyanoacetaldehyde, precursor for malononitrile. |
| NCCN | Cyanogen | purine precursors, precursor for cyanate[31] | If nitrosated, for purine synthesis |
| HCHO | Formaldehyde | carbohydrate precursor[4b] | Reservoir with $SO_2$[9] |
| $HOCH_2CHO$ | Glycolaldehyde | carbohydrate precursor[4a, 4b] | Reservoir with $SO_2$[9] |
| $H_2NCONH_2$ | urea | nucleobase precursor[32] | Phosphorylation[11b, 33] |
| $HCONH_2$ | formamide | nucleobase precursor[34] | Matrix for anhydrous reactions[35] |
| $NH_3$ | ammonia | components of nucleobases | Activate trimetaphosphate[10] |

**Table 2**. Inorganic species possibly involved in prebiotic RNA synthesis models

| Class | Example* | Role in RNA formation |
|---|---|---|
| Phosphate | Apatite, whitlockite | Sources of phosphate[33, 36] Catalyst[4a] |
| Borate | Colemanite | Form pentose reservoir, enolization catalyst/moderator[4b, 12, 37] |
| Borophosphate | Lüneburgite | Form ribonucleoside phosphates & diphosphates[13a, 13c] |
| Volcanic gases | $SO_2$ | Form reservoirs of lower carbohydrates[9] |
| Trimetaphosphate | $Na_3(PO_3)_3$ | Phosphorylation, direct glycosidic bond formation[10]] |
| Silicate | Opal | Binds/stabilizes RNA oligos[38] |
| Carbonate | Calcite | Binds oligomeric RNA[39] |
| Molybdate | Powellite | Catalyzes carbohydrate rearrangement[40] |
| Phosphide | Schreibersite | Kinetically facile P-O-C bonds formation[41] |

* not necessarily the prebiotically relevant mineral

### The oxidation/reduction states of organic species and minerals

The redox state of species in **Tables 1** and **2** is key to the review that follows. In organic chemistry, the terms "oxidized" and "reduced" refer to the ratio of hydrogen atoms to oxygen, nitrogen, and other "hetero" atoms attached to a carbon scaffold. Consider, for example, organic molecules containing just one carbon atom in their scaffold. Here, a redox scale begins with methane ($CH_4$) as the most reduced molecule, proceeds to methanol ($CH_3OH$), formaldehyde (HCHO), and formic acid (HCOOH), and ends at carbon dioxide ($CO_2$) as the most oxidized molecule.

The carbon atoms in RNA have oxidation states ranging (depending on how states are assigned) from $CH_3OH$ through $CO_2$. All of the nitrogen atoms in RNA are at the oxidation state of ammonia ($NH_3$) or a primary amine ($C\text{-}NH_2$). This range is within the range of oxidation states of the carbon and nitrogen atoms in HCN, $H_2NCN$, HCCCN, and the other primary nitrogen-containing RNA precursors. This allows RNA to conceptually form from these precursors with neither oxidation nor reduction. Experimentally, routes to the RNA nucleobases are also known from these precursors.[5, 42]

Geologists describe mineral oxidation states differently. Their descriptions reference an "oxygen fugacity" (an effective partial pressure of oxygen) and the relative oxidizability of atoms in a mineral. Thus, minerals invoked by various path-hypotheses, including phosphate,[10] borate,[4b, 12] phosphoborate,[13a, 13c] silicate, aluminosilicate, molybdate,[40] sulfite,[9] and sulfate (**Table 2**), all have abundant oxygen. Of these, sulfite ($SO_3^{2-}$), sulfate ($SO_4^{2-}$), and molybdate ($MoO_4^{2-}$) require environments that have an oxygen fugacity near or above the fayalite-magnetite-quartz equilibrium (FMQ, $Fe_2SiO_4$-$Fe_3O_4$-$SiO_2$) that typifies Earth's upper mantle today.[43] Phosphate ($PO_4^{3-}$), borate ($BO_3^{3-}$), and silicate ($SiO_4^{4-}$) are also available in melts having an FQM fugacity, as well as in more reducing environments.

Phosphate is a component of RNA at this 5+ oxidation state. However, lower oxidation states (e.g. phosphite $PO_3^{2-}$) have been invoked as intermediates to create phosphorylated organics.[41] Iron redox geochemistry may have made reduced phosphorus available in the Hadean,[44] but low-oxidation phosphorus (effectively $P^0$) is also abundant in iron meteorites. The second is important in the impact scenario described below.

Phosphorus in its 5+ oxidation state is invoked in path-hypotheses that directly form nucleotides from heterocycles and ribose,[11b] indirect routes to secondary RNA precursors (**Fig. 2**),[4a] and in routes that generate nucleoside 5'-phosphates and diphosphates from nucleosides.[13c] Sulfur in its 4+ oxidation state ($SO_2$, bisulfite, and sulfite) allows the accumulation of substantial amounts of carbohydrate precursors in stable form.[9] Oxidized borate (boron in its +3 oxidation state) may productively guide reactions to form carbohydrates, nucleosides, and phosphorylated nucleosides.[4b, 12-13]

### Geological measurements suggest that the Hadean mantle had a FMQ -0.5 ± 2.3 oxidation state

The importance of redox states to both the organic chemistry and the mineralogy of the Hadean makes it important to review the evolution of the redox state of the early Earth. Many measurements, observations, and models constrain its timeline.

Drawing on the physics of planetary accretion, much of the original mass of Earth evidently came from material akin to enstatite chondrites and aubrite meteorites.[45] Still falling in relatively small numbers to Earth today, these meteorites have a highly reduced mineralogy, with 20%-30% metallic iron.[46] In the aggregate, this material is more reduced. Important to our comparison between mineralogy and organic chemistry, their reduced state is compatible with the reduced state of carbon and nitrogen in HCN, $H_2NCN$, HCCCN, and other primary RNA precursors, as well as RNA itself.

However, a mantle this reduced could not have lasted very long. In an Earth molten from impact energy and heated by radioactive decay of short-lived nuclides such as $^{26}Al$, metallic iron and its reducing power must have sunk rapidly to the core in Earth's molten form, under the force of gravity. This must have left behind a mantle having a redox state between FMQ and the "iron-wüstite" buffer.[47] Analysis of the ratio of $^{182}Hf/^{182}W$ isotopes and uranium and lead isotopes suggests that this differentiation occurred rapidly, within 50 million years.[48] While Theia's impact at ~4.51 Ga[49] added iron metal to the already-formed terrestrial core, the core thereafter was largely chemically "closed" to the upper mantle. Afterwards, the iron metal and the associated siderophiles from the original accretion no longer influenced the mantle and the crust, which thereafter were redox-buffered at a higher FMQ oxygen fugacity.[50]

Thus, accretion models from physics suggest that in less than ~50 million years after formation of the Moon, with following ferrous iron disproportionation[50] and atmospheric processes discussed below, the Earth's mantle had achieved a level of oxygen fugacity comparable to that of the modern Earth. This made the minerals listed in **Table 2** available for prebiotic chemists to use in their path-hypotheses. Throughout the Hadean, crustal silicon was likely present as silicate, phosphorus as phosphate, sulfur as sulfite or sulfate, and boron as borate. Even molybdate ($Mo^{6+}$, its highest normal oxidation state) can be present at FMQ fugacity in molten silicates,[43] a fact that surprises some.[18]

Experimental data from geology can assess the Hadean mantle redox state directly. For example, Trail and his coworkers analyzed cerium in Hadean zircons having uranium-lead dates as old as 4.36 Ga. Cerium exists in two oxidation states ($Ce^{3+}$ and $Ce^{4+}$), and laboratory experiments allow the $Ce^{3+}/Ce^{4+}$ ratio in a zircon crystal to serve as a proxy for the redox state of the melt from which it was crystallized. These experiments show that Hadean zircons were evidently formed in silicate melts with average redox environments of FMQ -0.5 ± 2.3, comparable to the oxidation state of Earth's modern mantle.[17] Further, inclusions in these zircons as well as $^{18}O/^{16}O$ ratios suggest that at least some continental crust may have been present > 4.3 Ga,[51] and that liquid water was likely present on Earth's surface $\geq$ 4.3 Ga.[52]

Controversially, but relevant to biology, isotopically light carbon ($^{13}C$ depleted relative to $^{12}C$), possibly indicating photoautotrophic fixation of $CO_2$, is reported in a single zircon that dates at 4.1 Ga.[53] If the carbon inclusions have the same 4.1 Ga antiquity as the uranium-lead system that dates the zircon, and if no non-biological process generates such light carbon, Earth may have had a biosphere already at 4.1 Ga.

The interpretation of the origin of any inclusions in zircons surviving from the Hadean is hotly disputed, in particular whether they are primary or arise via later contamination. This is especially so with the zircon with the reported carbon inclusion.[54] In part, the controversy arises because it is a single example, not supported by a series of increasingly younger zircons containing increasingly lighter carbon. Nor is the geohistory of $^{12}C/^{13}C$ ratios well defined. The next oldest (and similarly controversial)[55] evidence for life based on $^{12}C/^{13}C$ ratios in ancient rocks is found in 3.85 Ga rocks from the Akilia association in West Greenland.[56] Biogenically light carbon preserved in rocks that are likely sedimentary in origin is secure only ca. 150 million years later.[57] Non-biological processes might generate light carbon.[58]

### An atmosphere in redox equilibrium with an FMQ Hadean mantle has little reducing gas

After the impact by Theia that generated the Moon, and after the iron core became largely isolated from the upper mantle, the Earth cooled rapidly (in less than 10 million years).[59] Thereafter, it is widely accepted that the redox state of the Hadean atmosphere was largely determined by the redox state of volatiles outgassing from the mantle.[60]

Assuming a mantle having an FMQ redox state, $CO_2$ was likely the principal form of carbon delivered to the atmosphere by outgassing, along with some CO but only small amounts of $CH_4$.[61] The principal form of sulfur delivered to the atmosphere was likely $SO_2$ (which can disproportionate to give sulfate and elemental sulfur), with some $H_2S$. The principal form of nitrogen was likely $N_2$, with a little $NH_3$. The principal form of hydrogen was likely $H_2O$, with little $H_2$. These are often referred to as "oxidatively neutral" **(**$CO_2$ + $N_2$ + $H_2O$)[62] or "weakly reduced" (if they contain some $H_2$, $CH_4$, or CO) atmospheres.[63]

### An atmosphere with little reducing gas cannot generate much reduced primary RNA precursors

This leads to an apparent paradox. An oxidatively neutral Hadean atmosphere is much more oxidized than HCN, $H_2NCN$, NCCN, and HCCCN. It is also more oxidized than HCHO and glycolaldehyde, commonly invoked as precursors for RNA ribose. Thus, an oxidatively neutral atmosphere is not a productive source of any of these nitrogenous precursors, even though each of the path-hypotheses in **Fig. 1**-**4** requires at least some of them.

We start with HCHO (formaldehyde), which is prominently featured in many path-hypotheses as a precursor for ribose, the "R" in RNA.[4b] Experiments and models show that the rate of production of HCHO in standard atmospheres is largely independent of the redox composition of those atmospheres (**Fig. 5**).[63a] HCHO was thus likely generated in large amounts in many atmospheres, even in redox neutral atmospheres with $CH_4$:$CO_2$ ratios near zero.[64] These ratios are currently considered to be the most likely in Earth's atmosphere at the relevant time in the Hadean.

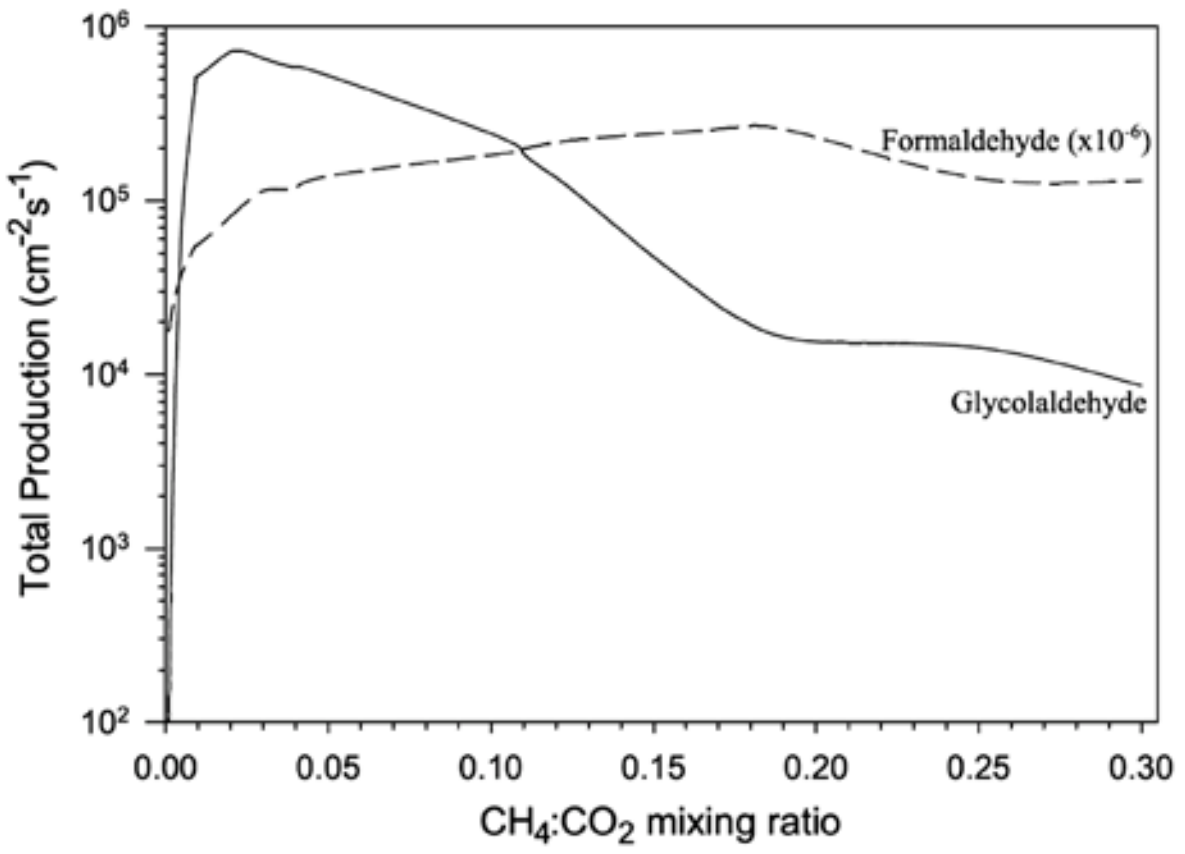

**Figure 5**. Atmospheric photochemistry models for the production of formaldehyde (HCHO molecules per square centimeter surface area, dashed line) and glycolaldehyde ($HOCH_2CHO$, solid line) for a range of $CH_4$:$CO_2$ ratios. Here, the HCHO values are scaled a million fold higher relative to the numbers on the *y* axis; thus, at a $CH_4$:$CO_2$ ratio of 0.12, where the two lines cross, the amount of HCHO formed (which is ∼2 × $10^{11}$ molecules $cm^{-2}s^{-1}$) is one million times larger than the amount of glycolaldehyde formed. If HCHO precipitated undisturbed on the surface, the accumulation after a year would be ca. 0.1 mole, or 3 g/$m^2$ per year. Over a million years, the accumulated HCHO (if unreacted, perhaps in the form of paraformaldehyde) would be 3 x $10^6$ grams/$cm^2$, approximately ~2 meters deep. From Harman et al. (2013), with permission.[63a]

In contrast, the production of glycolaldehyde in an atmosphere depends much more strongly on its redox state. Glycolaldehyde is not formed in detectable amounts in atmospheres lacking $CH_4$ (**Fig. 5**).[63a] Even in atmospheres containing substantial amounts of $CH_4$, glycolaldehyde is formed only in very small amounts. Thus, in a mixture of gases with a $CH_4$:$CO_2$ ratio of 0.02, glycolaldehyde is formed at one part per million of HCHO (**Fig. 5**). This may be sufficient for a catalytic role for glycolaldehyde in the formation of pentoses, including ribose.[4b] However, this rate of formation is not alone sufficient for glycolaldehyde to be a stoichiometric contributor in a high probability scenario for forming RNA.[4a, 63a]

The challenge is increasingly severe with HCN, $H_2NCN$, HCCCCN, and NCCN as nitrogen-containing precursors of the RNA nucleobases. None of these enter into any path-hypothesis as a catalyst. All are required mole-for-mole in the synthesis of RNA building blocks, as they are for most "grandfather's axe" heterocycles. All of the primary precursors and heterocycles invoked in these path-hypotheses have nitrogen atoms at a low level of oxidation.

Thus, once it reached nearly its present mass, and most likely before it suffered its last sterilizing impact, Earth's mantle was likely to have never been sufficiently reduced to be comparable in redox potential with an atmosphere productive in reduced RNA precursors. This is especially true for primary nitrogen precursors (HCCCN, $H_2NCN$, and others) having higher molecular weights,[50, 65] which require more "reducing power" than that needed to make the smaller HCN.[66] Indeed, considering the photodissociation of $NH_3$ and $CH_4$ followed by the escape of $H_2$ to space, the atmosphere would generally tend to be more oxidized than the gases that supplied it from below. For this reason, all of the path-hypotheses considered here (**Figs**. **1-4**) for the abiological formation of precursors for RNA are improbable to the extent that they rely on the formation of primary precursors in an oxidatively neutral Hadean atmosphere.

**Ways of getting reducing power into the Hadean atmosphere**

Many efforts seeking to develop path-hypotheses overlook this. Others look to a late delivery of reduced organic materials from outside the Earth, perhaps via comets,[67] the formation of reduced precursors by impact energy,[68] or the influx of energetic particles from a young Sun.[69] Still others look to a flux of high-energy photons to react with cyanide coordinated to transition metals to generate mixtures of reduced species, with reducing power ultimately coming from $SO_2$ (**Fig. 4**).[18, 22]

Ultraviolet irradiation, lightning, impact plumes, meteor ablation trails, and silent electrical discharge can indeed generate reduced species from $CO_2$, $N_2$, and $H_2O$ in redox neutral atmosphere.[70] However, the yields of these reduced precursors depend on the details of the redox state of the atmosphere.[71]c Further, the stoichiometry of such processes in a redox-neutral atmosphere produces a matched oxidized species for every reduced species. This makes such reactions neutral overall in oxidized and reduced species, regardless of the source of energy that generates them. Further, since $H_2$ is the easiest gas to gravitationally escape from the Earth's atmosphere, the long-term trend in an atmosphere is largely oxidizing.

Of course, reducing power can come from the mantle itself, even at FMQ. For example, $H_2$ can be produced by serpentinization, the aqueous corrosion of iron silicates like fayalite;[72] that $H_2$ can lead to the formation of $CH_4$.d However, the amounts of these (t most tens of parts per million[75]) seem too small to make the Hadean atmosphere richly productive in HCN, and unable to support at all useful atmospheric production of higher molecular weight reduced nitrogen compounds (**Fig. 6**).[66, 76] Smaller amounts of HCN might be captured as complexes with various metals; ferrocyanide has been long discussed,[21, 22] but other metals might be considered.[77] The value of these as sources of prebiotic HCN remains unclear.

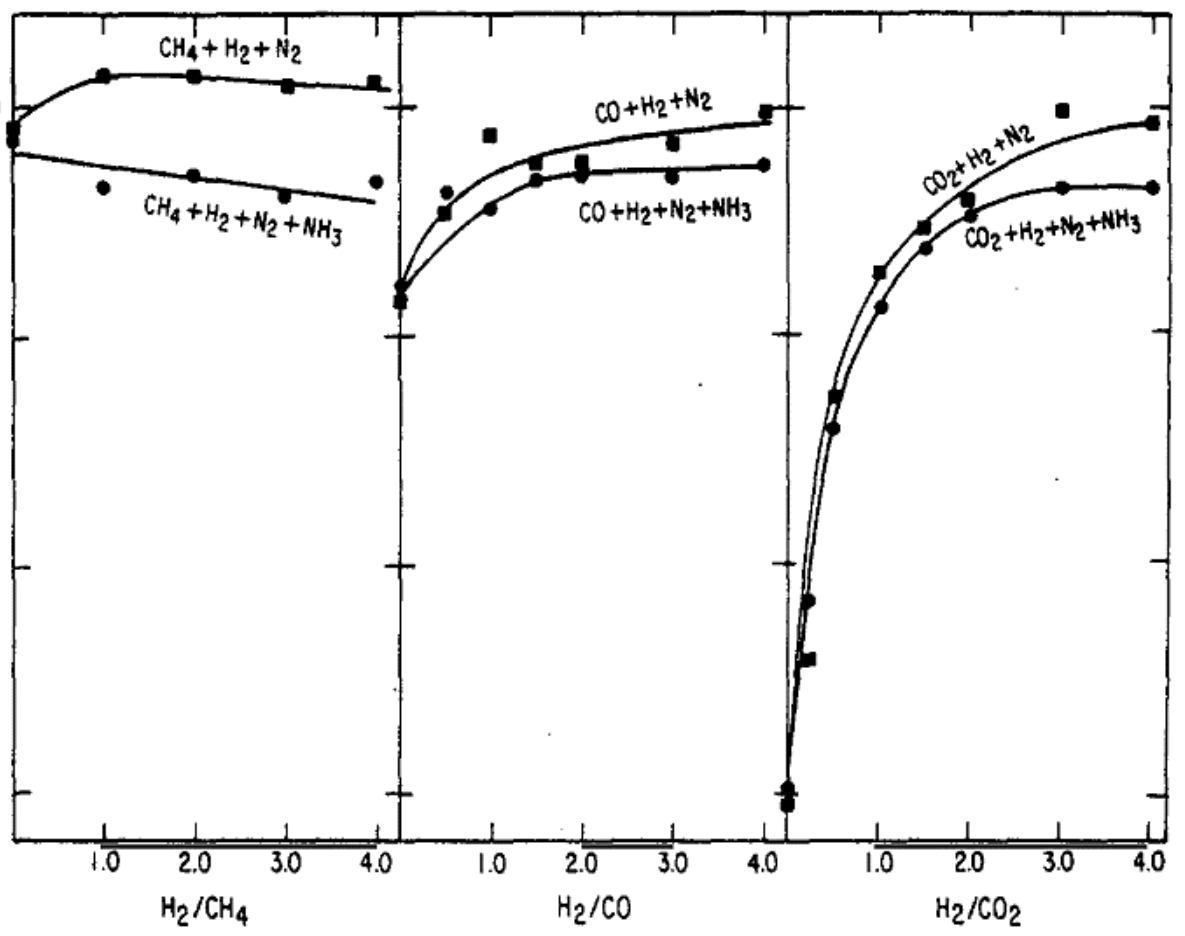

**Figure 6**. Yield of hydrogen cyanide (HCN) from laboratory electrical discharge experiments in various mixtures of gases. Reduced carbon (CO, $CH_4$) is required for substantial yields of HCN. Thus, prebiotic syntheses of nucleobases that require large amounts of HCN cannot proceed in the standard consensus atmosphere (right panel). From Miller & Schlesinger (1983), with permission.[76]

**Impacts have the potential to deliver large amounts of reducing power**

This paradox might be resolved by emerging models for the impact history of early Earth. Here, the relevant observations begin with an analysis of the siderophilic ("iron loving") elements in Earth's mantle and the Moon. Siderophiles include osmium, iridium, platinum, rhenium, ruthenium, rhodium, and gold; all are extracted into iron during metal-silicate equilibration. Thus, virtually all siderophiles that arrived before the closure of the Earth's iron core would have sunk to the core with the metallic iron that now forms the core. All of the initially accreted siderophiles would have been lost from the mantle.[78]

Nevertheless, it is evident that the Earth's crust-mantle has gold and other siderophiles in what is called a "veneer". Their concentrations are much higher than expected if they had been delivered before core closure.[79] The best literature estimates put these at 0.3-0.8 wt% of the Earth (e.g. Day et al. (2007), Willbold

c Impactors have also been proposed to be *direct* sources of primary and/or secondary organic precursors of RNA. For example, the adenine in carbonaceous chondrites might have been incorporated directly into RNA.[67] We do not explicitly consider these sources due to the relative scarcity of carbonaceous chondrites in Earth's accretion history.[71]

d As reviewed in Kasting & Catling (2003),[73] electrical discharge through $CO_2$-$N_2$ mixtures produces NO rather than HCN.[74] The amount of HCN formed in the upper atmosphere via the photochemical ionization of $N_2$, with the amount of HCN formed depending on the amount of $CH_4$ present. Kasting & Catling (2003) estimate that serpentinization may have produced an abiotic flux of $10^{13}$ moles $CH_4$ per year, with the atmospheric steady state depending on photochemical loss.[73]

et al. (2015), Day et al., (2016)).[78, 80] These elements are believed to have arrived later, after the core became closed, via impacts of bodies too small to re-melt the Earth entirely and re-differentiate its constituents.[81]

As extremes, the amounts of siderophiles observed in the mantle's veneer may have come from one larger impactor[82] or from many, many small impactors.[83] These alternatives might be distinguished by the fact that Earth's veneer is 1950 ± 650 times larger than the Moon's.[78, 81a, 84] This ratio is much larger than the relative gravitational cross-sections of the Earth and the Moon. The Earth, by these measurements, has too much veneer, while the Moon has too little.

The Moon could have avoided a size-commensurate siderophilic veneer by the statistics of small numbers. If only one (or a very few) large impactors delivered all (or most) of that veneer, one or the other body would have collected that veneer, and the other would lack it. The Earth was more likely than the Moon to have received that impact. Under this scenario, the amount of the veneer determines the size of the (single) impactor that delivered it. For example, a $10^{25}$ kg impactor (such as Theia) is too large to account for the observed veneer. Further, a second Theia impact energy would have re-melted the Earth, causing reformation of the core; any siderophiles that it did deliver would have been lost to that core, leaving behind no veneer.[85]

Conversely, a single $10^{21}$ kg impactor (about the size of the modern asteroid Ceres, ~1000 km diameter) would have been too small to deliver the observed amount of late veneer in a single impact. Approximately 80 of these $10^{21}$ kg impactors would be required to explain the amount of the observed terran veneer; this number of impactors would have led to an Earth-Moon distribution of the veneer in a ratio close to their gravitational cross-sections, which is not what is observed.

Rather, a $10^{23}$ kg impactor (ca. today's Moon, ~1% Earth mass, ~2700 km diameter) fits the geochemical constraints. Called Moneta by some (to reflect its hypothesized delivery of metals important in money), a $10^{23}$ kg impactor is large enough to deliver in a single impact the required amount of the siderophiles in the late veneer. However, the hypothesized Moneta was not so large as to cause the re-formation of the Earth's core, into which the veneer would have been lost by metal-silicate equilibration.[86] Further, Moneta was large enough to have had its own iron core.

**A Moneta-sized impact would have opened a window for forming reduced RNA precursors**

If a $10^{23}$ kg Moneta-sized impactor is concluded to be the way that most of the veneer was delivered to Earth (and not the Moon), this conclusion can also resolve the apparent paradox arising from the unavailability of the primary precursors for RNA under a redox-neutral Hadean atmosphere. A $10^{23}$ kg impactor would have had its own iron core as a result of its own independent accretion and differentiation history. In its most probable oblique impact trajectory (45°), as well as many others, that iron core would have shattered, delivering molten iron droplets (>1820 K) as a metallic hail to the Earth's atmosphere and crust, rather than delivering it to the Earth's pre-formed core.[87] Further, this event is large enough to have re-set most radioisotopic clocks on Earth.[88]

Molten iron from an impactor whose size, by itself, accounts for the amount of the terran veneer, unavoidably would have reduced much of the accessible water to dihydrogen ($H_2$). The iron core from a $10^{23}$ kg Moneta impactor could have reduced up to three ocean masses, delivering as much as 90 bars of $H_2$ to the atmosphere. This atmosphere would then have had a redox state fully compatible with the formation of HCN, HCCCN, and the other reduced primary RNA precursors cited above (Parkos et al. (2018) and references therein).[89] Metallic iron also reduces $CO_2$ to CO and methane, and reduces $N_2$ to $NH_3$. The resulting atmosphere would then have become a productive source of large amounts of HCN, HCCCN, $H_2NCN_2$, glycolaldehyde, and other primary precursors that are inaccessible in the standard Hadean atmosphere.[61]

For example, in certain mixtures of reduced $CH_4$ and $N_2$, cyanoacetylene (HCCCN) is produced in a 9% yield,[30a] high by the standards of prebiotic chemistry. HCCCN is an efficient precursor for the RNA nucleobases cytosine and uracil. Analogous yields of HCN are produced by electrical discharge in atmospheres with reduced $CH_4$ or CO in the presence of $N_2$ or $NH_3$ (**Fig. 6**).[76, 90] Electrical discharge through mixtures of $N_2$, $H_2$, and $CO_2$ gases can give (for example, in a $H_2$:$CO_2$ 3:1 ratio) large amounts (10-15%) of HCN;[91] HCN is a proposed precursor for adenine and other nucleobase fragments.

While its impact on the redox potential of the atmosphere would have been substantial, the iron from the $10^{23}$ impactor would not have substantially altered the redox potential of the mantle. Genda et al. (2017)[87] estimate the amount of FeO formed from the impactor's iron core at ~0.6 wt% of the FeO already in the Earth's mantle. This can be compared to the amount of FeO in today's mantle (~8 wt%).[92] Thus, oxidized minerals would continue to be available from the Hadean mantle, even though Moneta would have most likely created a temporary magma ocean on the surface, and delivered a pulse of reduced transition metals and phosphorus to the surface and the crust. This is a resource for the phosphorylation of RNA building blocks in various path-hypotheses that invoke phosphorus at one of its lower oxidation states.[41, 93]

**These reduced organics formed in the atmosphere would have precipitated**

The reduced organic molecules formed in the post-impact reducing atmosphere would not have remained there long. Even HCHO, nominally a gas, would have dissolved in atmospheric water droplets as its hydrate ($HOCH_2OH$), unavoidably forming a bisulfite adduct ($HOCH_2SO_3H$) with volcanic $SO_2$, which was likely produced in comparable amounts by outgassing of the FMQ mantle. These droplets would have rained HCHO, glycolaldehyde, and their bisulfite adducts to the surface.[9] Likewise, HCN, also nominally a gas, would have dissolved in atmospheric water droplets and either rained to the surface directly or as its hydrolysis products, formamide and ammonium formate. Interestingly, if the pH of these droplets were controlled by volcanic $SO_2$ (pH ~4), the half-life for HCN would have been maximal (~$10^2$ years at 60 °C),[94] and the amount of HCN delivered the largest.

Similar precipitation, before or after hydrolysis, would be seen for the other reduced organic precursors made in the post-impact reducing atmosphere. These include $H_2NCN$ (hydrolyzing first to urea, also valuable, and then to ammonium carbonate) and HCCCN (hydrolyzing first to $NCCH_2CHO$, also valuable). Cyanogen (NCCN) hydrolyzes first to $NCCONH_2$ in weak acidic solution;[95] in base, NCCN hydrolyzes first to HCN and HOCN;[31] various other products are produced by further hydrolysis. All of these would have precipitated to the surface, together with their further hydrolysis products.

**Subaerial land that is intermittently dry is required**

But *to where* would they have precipitated? If the Hadean Earth was entirely covered by water, these reduced RNA precursors would have rained into a global ocean, where they would likely have been diluted into unproductively low concentrations (but see Russell & Hall (1997), Russell et al. (2010)).[96] Absent some (as yet unknown) mechanism for concentrating and stabilizing them in the ocean, most would eventually hydrolyze. Thus, all of the path-hypotheses considered here require the capture of the precipitating primary reduced RNA precursors onto sub-aerial surfaces, only intermittently exposed to water.

Of course, immediately after the magma ocean created by the Moneta impact had solidified, exposed surface is likely, since much of the terran water would have been reduced and much of the rest turned to steam. Relatively little is known about the transformation of these primary RNA precursors precipitating onto hot rock beneath a steam-rich atmosphere. However, RNA and most of its advanced precursors would not likely be stable.

Even after the magma surface cooled, a dipole-induced greenhouse arising from various atmospheric collision–induced absorption processes (e.g. $H_2$-$H_2$, $H_2$-$N_2$, $H_2$-$CH_4$, $H_2$-$CO_2$)[97] may have kept the surface unproductively hot for millions of years. These factors would delay the opening of the "window of opportunity" for RNA formation, even given an atmosphere able to produce its primary precursors.

As time passed, however, Earth's surface temperature dropped and its water inventory condensed to form larger oceans; here, it is not clear how much subaerial land remained. Models and speculations for the amount of continental crust (not the same as subaerial land) at 4 Ga range widely (**Fig. 7**), from zero to the modern level. This range is the largest uncertainty in this discussion, as the probability of RNA forming is presumed to be an increasing function of the amount of land on which it can form from precipitating precursors. Authors who see the Hadean Earth as a "waterworld" require that life emerge on the small acreage of ancient volcanoes, analogs modern Hawai'i,[15] or geologically active environments analogous to modern Iceland.

However, oxygen isotopic data and the presence of inclusion assemblages in zircons of the relevant age suggest the existence of at least some land as a place for sub-aerial weathering reactions to form clay-rich sediments. These perhaps suggest a total sub-aerial surface that might be estimated to have been similar to the surface of modern Australia.[98] Authors accepting this view look to isolated ponds and lakes on land intermittently exposed to water (as in modern Death Valley) as places where RNA might assemble.[4b] Again, these considerations are not changed depending on what path-hypothesis one prefers from **Figs. 1-4**.

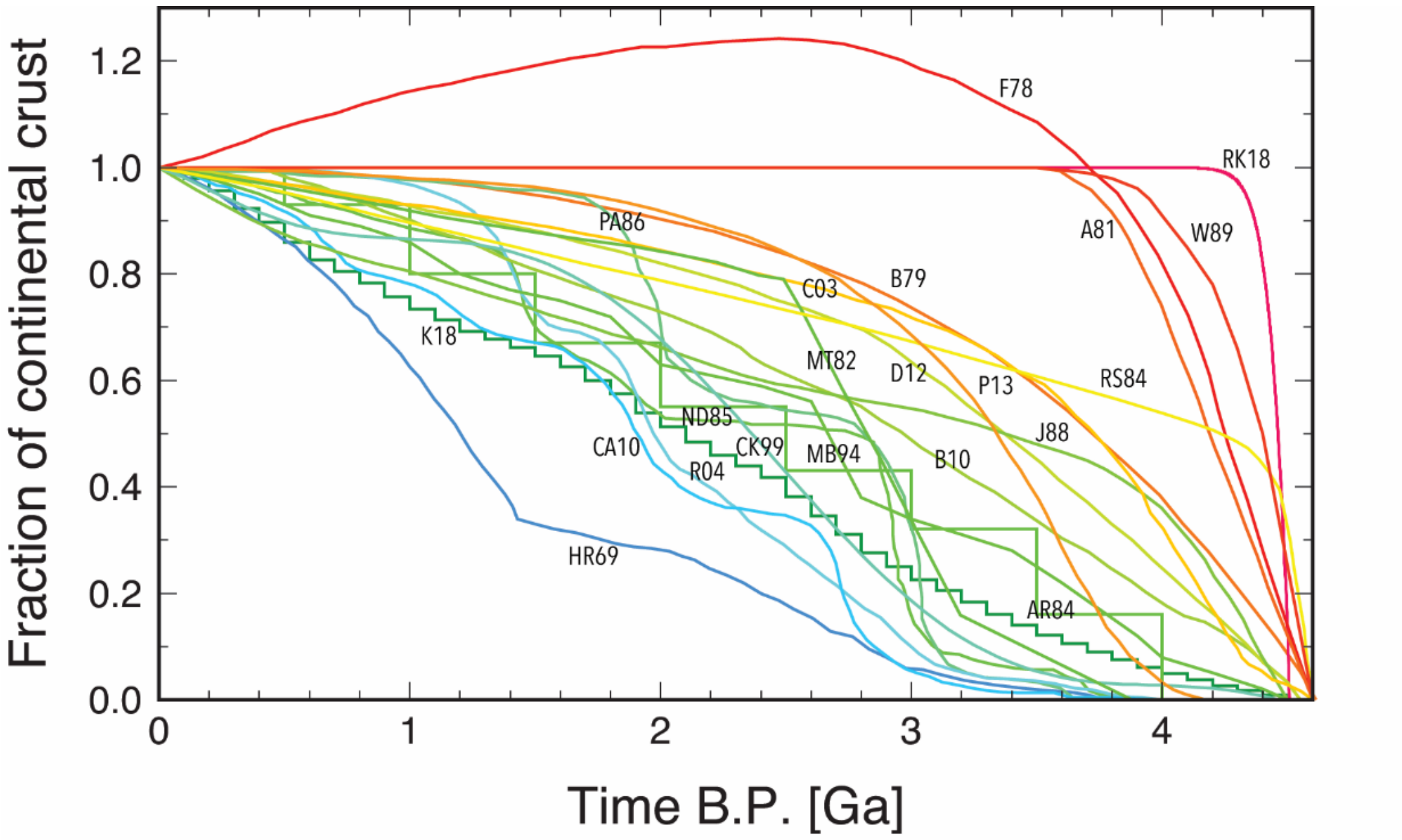

**Figure 7.** Various models for the growth of continental crust, from Korenaga (2018), with permission.[98] While models differ widely, all have the amount of crust increasing over time. The spectrally red models all provide abundant crust at relevant times in the Hadean, and do not drive the date when RNA formation was most probable. However, models represented by yellow and green curves drive that date towards the end of the window of atmospheric productivity, if dry land is required for RNA formation. The models represented by blue curves, of course, preclude any land-based RNA formation prior to 3.8 Ga. Analyses of mica inclusions and oxygen isotope data in Hadean zircons suggest that the Hadean did have a small volume of crust with a composition similar to that of the modern crust.[98a, 100] However, the continental area (and, by inference, the amount of sub-aerial land) should not be confused with the volume of continental crust. Even if the Hadean continental crust volume were the same as today, most of the crust was likely to be underwater because the oceans were likely more voluminous then.[101]

In addition to capturing reservoirs of secondary RNA precursors, sub-aerial dry land is needed in path-hypotheses for several late steps in abiological RNA synthesis. For example, in the path-hypothesis summarized in **Fig. 1**, three steps require dryness: (a) the formation of the glycosidic bond of nucleosides, (b) the formation of nucleoside phosphates and diphosphates, and (c) the formation of oligomeric RNA.

This sets up two competing factors influencing a view of when RNA most probably emerged. The productivity of the reducing atmosphere would have been highest soon after the impact. Over time, $H_2$ (and its reducing power) would have been lost by escape to the cosmos under UV irradiation from the early Sun.[102] Genda et al. (2017) estimate that ~200 million years of hydrodynamic escape would be sufficient to effectively eliminate atmospheric $H_2$.[59, 81e] Emission of relatively oxidized gases from the mantle is expected to have further acted to restore the consensus redox-neutral Hadean atmosphere. Over time, the rate of production of precipitating RNA precursors would have dropped. The cartoon in **Fig. 8** models this as a single exponential with a 40 Myr half-life. After (arbitrarily) three half-lives, the atmosphere would have returned to a redox state unproductive for the most probably useful formation of RNA precursors. This would close the window of opportunity for RNA formation on sub-aerial surfaces.

Again, this constraint is robust for alternative path-hypotheses. For example, phosphate esters might form by regioselective phosphorylation by magnesium borophosphate mineral (e.g. lüneburgite), by apatite in a slightly acidic eutectic,[13c, 33] by activation of cyclic trimetaphosphate by ammonia,[10] or by phosphite arising from the corrosion of the mineral schreibersite.[41] All require dryness, if only intermittent.

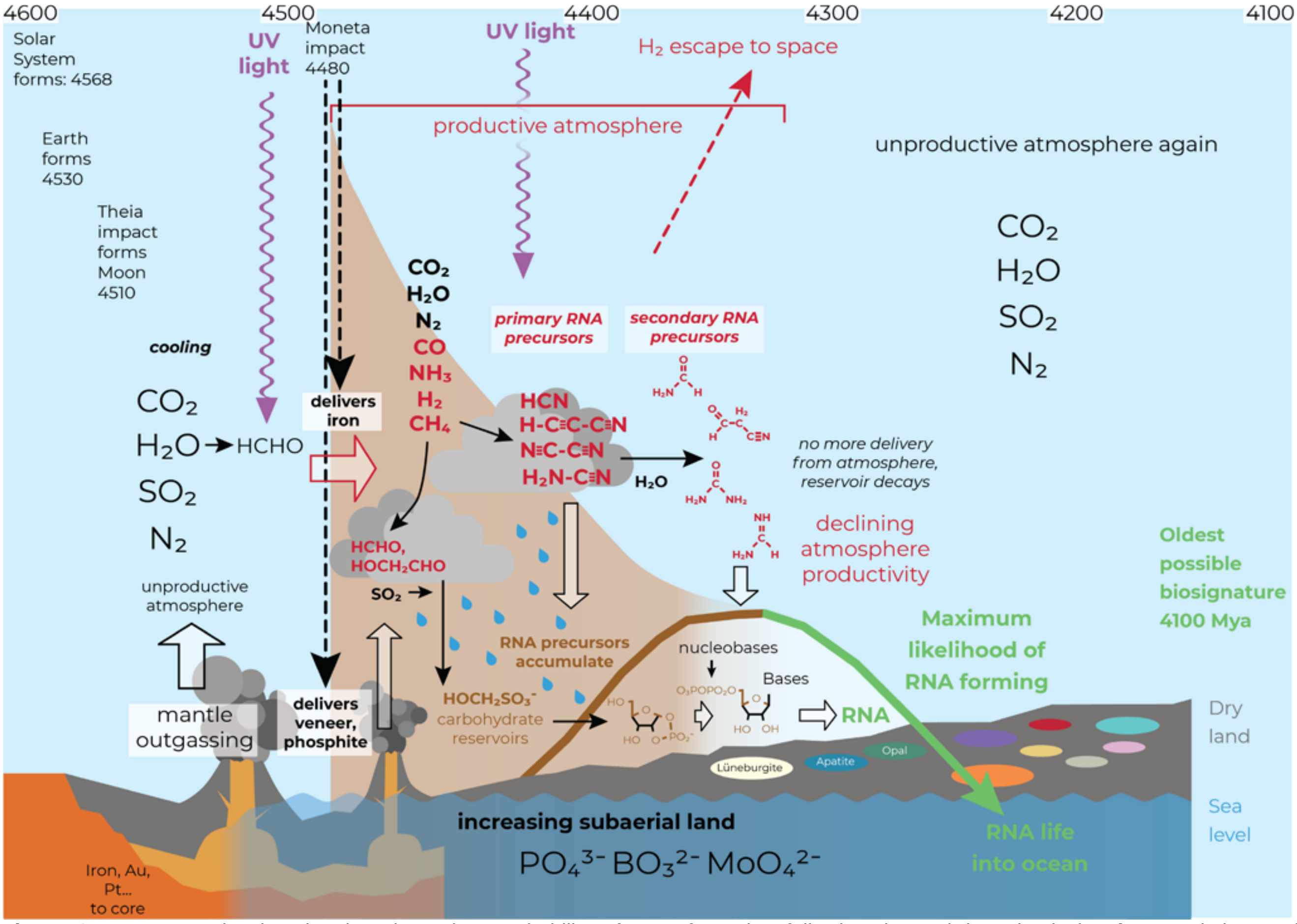

**Figure 8**. A cartoon showing the time-dependent probability of RNA formation following the path-hypothesis in **Fig. 1**, relative to the hypothesized Moneta $10^{25}$ kg impact ($t_0$ = 0) that could have delivered the late veneer in the amounts observed. Top time scale in Ma units. Iron from this impactor would have converted an unproductive atmosphere (pale blue, dominated by species shown in black) into a productive reducing atmosphere (brown), which allowed the formation of reduced organic molecules (red primary RNA precursors), which (in part) hydrolyzed to give secondary RNA precursors (also red); some of these were captured and stabilized by $SO_2$ emerging from the relatively oxidized mantle to give carbohydrate reservoirs (brown). All precipitated onto sub-aerial land, shown to be increasing over time. The reducing productivity of the atmosphere decayed over time via escape of $H_2$ (red arrow, here modeled with a half-life of 40 Myr) and outgassing from an FMQ mantle. This precipitation allowed the accumulation of primary, secondary and more advanced RNA precursors on the surface accessible to oxidized mineral species from a FMQ mantle (some are shown in ellipses) to generate nucleosides, phosphorylated nucleosides, and oligomeric RNA that binds silicate species. The physics of $H_2$ loss makes the variance on the window of opportunity is surprisingly small (± 100 Myr). If $t_0$ = 4.48 Ga,[71b] RNA formation may (non-analytically) be most probable at 4.36 ± 0.10 Ga before present, three half-lives into the atmosphere decay. Different rate assumptions, and different assumptions about the amount of sub-aerial land, move the time where RNA formation is seen to be most probable, but only modestly. Continued rain-out after the start of RNA-based Darwinian evolution would provide sustenance for this first life on Earth.

### This defines a window of opportunity for RNA formation

The amount of reduced organics delivered in this scenario from the atmosphere after the magma ocean had cooled would have been substantial. Estimates of the productivity of Hadean atmospheres vary.[63a, 64] However, as with the calculation for HCHO above, meters of depth of organic material on dry surface areas would have been delivered. Absent loss, organic RNA precursors would have accumulated.

Any primary and secondary RNA precursors that accumulated presented an opportunity for advanced RNA precursors to form.

These include intermediates shown in **Figs. 1–4**, as wells as "grandfather's axe" precursors and others not shown. We assume, perhaps reasonably, that the probability of RNA forming is an increasing function of the concentrations of precursors accumulated in a particular region. This assumption is again largely independent of which path-hypotheses one favors.

Here, Titan offers an imperfect model. On Titan, organic species formed via photochemistry high in its reducing atmosphere precipitate to and accumulate on the surface; their depth is uncertain, but it appears to be many meters. The organic haze arising from a relatively dry upper level (hydroxyl radicals from photolysis of water inhibit haze formation) prevents most photochemistry from occurring in the lower atmosphere or on the surface, where they are frozen at ~95 K and available for inspection by future NASA missions.[103] However, if the temperature were warmer (a "warm Titan"), the precipitated organics would react by standard chemistry. Those advocating the RNA-First hypothesis for the origin of Darwinian evolution hope that their reaction occurs via a proposed path-hypotheses, perhaps one summarized in **Figs. 1**-**4**.

Of course, the surface, even its subaerial part, would not have been uniform, one of many contrasts with Titan. Regions near volcanoes would have remained hot. Other regions would have been cold, perhaps below 0 °C, depending on the view of the brightness of the Sun at the relevant time. As another difference, Titan's cold atmosphere has relatively little atmospheric water; hydroxyl radicals from atmospheric water are believed to influence the distribution of atmosphere-generated organics.

### The window of opportunity for RNA formation would have been transient

The useful RNA precursors would not have continued to accumulate forever, especially if the Earth had large amounts of water. The simplest process for loss would be run-off from semi-dry land into a diluting ocean as rain fell. Erosion of the accumulated material by such meteorology would occur with a time scale of years.

With similar time scales, RNA precursors are subject to decay, first via hydrolysis. Some hydrolysis would be productive, such as the hydrolysis of cytosine to uracil;[104] both cytosine and uracil have prebiotic value as advanced RNA precursors. However, other hydrolytic processes do not have obvious prebiotic value, such as the hydrolysis of useful urea (**Table 1**) into largely useless carbonate. Even ferrocyanide, a potential reservoir for HCN, decomposes.[21]

Further loss involves self-reaction to form "tars", complex mixtures of organics believed not to be useful to initiate life (but see Gutenberg et al. (2017)).[105] Glycolaldehyde, glyceraldehyde, and other carbonyl compounds react with themselves and each other; they also react with ammonia and other species to form a range of products that seem to be prebiotically useless. While borate minerals and sulfite adducts stabilize these,[4b, 9, 37] this stabilization is not absolute. Further, while organic hazes would have prevented destructive photochemistry from occurring at the surface under a very productive reducing atmosphere, UV would have become available as the atmosphere returned to the unproductive oxidation-neutral state. Surface UV could then have destroyed many useful accumulated species. For example, ammonia is photo-oxidized.[106]

The amounts of accumulated organic molecules available for the prebiological synthesis of RNA precursors at all levels under the models shown in **Figs. 1-4** would therefore have reached a steady state. In this steady state, their rates of loss would match their rates of formation. As atmospheric productivity decreases, the steady state level of these RNA precursors would decrease. So would the probability of forming RNA, which we assume is an increasing function of the steady state levels of RNA precursors.

### Atmospheric productivity is likely required *after* RNA-based Darwinian evolution begins

The larger the impact, the longer one must wait for the surface to become hospitable for RNA formation, but also the longer the time before the atmosphere ceases to be productive of RNA precursors. Since the time constant for surface cooling is faster than the time constant for the loss of the reducing power of a productive atmosphere after an impact, the bigger the impact, the longer the time that the window of opportunity is likely to be open.

The length of the time that the window of opportunity is open is not only relevant to the probability of RNA forming, but also relevant to the time that RNA Darwinism could rely on "food from the sky" before it needed to evolve enzymes that catalyze the synthesis of its own food. This means that the probability of the *origin* of life under an RNA-First model might be transformed into a probability of origin *and survival* of life under that model. It will be ultimately interesting to ask whether the ribosome emerged before or after the atmosphere lost its productivity.

Thus, if the impact of Moneta created the last planet-wide sterilization, its window of opportunity for the origin and persistence of an RNA world would perhaps be the one most probable for the prebiotic synthesis of RNA. If a subsequent impactor too small to leave a signature in the veneer (perhaps partially) re-sterilized the planet, then the RNA world would need to have (at least partially) started over. With this smaller impactor, the window of opportunity would have opened a bit sooner, but that window for the origin and persistence of an RNA world would have closed much sooner. Thus, a later, smaller impactor is (non-analytically) expected to have created an integrated probability of origin and persistence smaller than a Moneta $10^{23}$ kg impactor.

These considerations have implications for life on exoplanets. A model where impact flux decreases over time (Theia, which resets everything, then Moneta, which is sterilizing but gives a large window of opportunity for RNA formation, and then possibly non-sterilizing Vesta) is grounded in observations of our Solar System (compare, for example, the two last lunar impact basins, Imbrium and Orientale). For extrasolar systems, impact history would be different. However, if later, smaller impactors that create shorter windows are more frequently sterilizing than recent literature suggests, then the probability of life originating on other planets decreases. Indeed, any model for the origin of life that requires contingencies (including impacts) universally lowers the probability of life arising.

### Balancing factors for a view of the most probable time for RNA formation after Moneta's impact

These factors influence the time when RNA formation might be viewed as having the highest aggregate probability, once Moneta's impact made the atmosphere productive for the formation of reduced nitrogenous compounds: (i) The productivity of the atmosphere as a decreasing function of time; (ii) the steady state concentrations of organics, decreasing as their rate of their replenishment decreases; and (iii) the amount of sub-aerial land on which that steady state might accumulate, perhaps an increasing function of time. This combination of factors is captured by the cartoon in **Fig. 8** for the path-hypothesis shown in **Fig. 1**. Here, a maximum likelihood for RNA formation is estimated non-analytically to occur at about 120 ± 100 million years after the Moneta impact ($t_0$), under a model where the half-life for atmosphere productivity decay is 40 Myr.

The range in the maximum probability estimate is earlier or later depending on uncertainties in the time-dependent strength of Solar radiation, the rates of loss of reducing power through outgassing of oxidized volatiles from the mantle, and/or the magnitude of atmosphere-surface interactions. **Table 3** lists a variety of factors that influence the time that the window of opportunity for RNA formation is open. The range in the probability estimate also reflects uncertainties in the amounts of sub-aerial land. Each of these are complex functions of the environment, and do not lend themselves to analysis.

**Table 3.** Time-dependent competing factors constraining window of opportunity for RNA formation

| Event | Constraint | Time dependence | RNA formation probability as a consequence of this factor |
|---|---|---|---|
| Surface temperature post impact | Organic mineral reservoir accumulation possible only <350 K | Rapid decreasing then stable, depending on atmosphere composition/solar activity | Increasing with surface cooling after impact, then constant after surface temperature reaches equilibrium |
| Reservoir formation | No RNA formation possible without precursors formed in a reducing atmosphere created by impact | Rate of reservoir formation decreases; accumulated reservoir increases over time; possibly steady state | Probability of RNA forming presumed to increase with amount/concentration of precursors in a reservoir in a locale, perhaps depending on underlying mineralogy |
| Reservoir destruction | Steady level of RNA precursors depends on relative rates of formation and destruction | Accumulated reservoir reaches a steady state early, then slowly decays | Probability of RNA forming depends on steady state reservoir levels, a complex function of reservoir formation |
| Sub-aerial land creation | Reservoir accumulation possible only on dry land* | Wide range of estimates | Probability of RNA forming increases over time as dry land increases |
| Reservoir loss | Wash into the ocean and are eventually subducted. | Reservoirs on land to achieve a steady state | Probability of RNA forming depends on amount at steady state |
| Redox potential of atmosphere | Required for reduced organic RNA precursors, especially nucleobases | $H_2$ escape; precursor production returns to zero over 200 My | Window of opportunity to form RNA to zero as atmosphere returns to consensus redox-neutral form |
| Volcanism | Release of $SO_2$ limiting step in formation of carbohydrate reservoirs | Little direct evidence. | Ability to accumulate $SO_2$ stabilized carbohydrate reservoirs decreases over time. |
| Solar intensity | Photon flux creates HCHO, HCN, other primary precursors | Main sequence stellar evolution | RNA production increases until haze clears as atmosphere returns to redox-neutral form, allowing photons to penetrate the surface. |

* Atmosphere-generated precursors that fall into a global ocean are presumed to have been diluted to a uselessly low concentration.

### The absolute date for the most likely formation of RNA under this impact scenario

These considerations offer a date range for the most likely formation of RNA (and, under the RNA-First model, for the emergence of life) relative to the Moneta impact that delivered the terran veneer and set $t_0$. To estimate an *absolute* date when the atmosphere became most productive, it is necessary to date the Moneta impact itself.

This is as difficult as for any date in the Hadean. The most resilient chronometers in the sampled worlds of the Solar System (the Earth, Moon, Mars, and Vesta) as well as meteorites from the asteroid belt, show no general resetting after ~4.45 Ga.[49a, 88b, 107] If the re-setting at or before 4.45 Ga corresponds to the veneer-forming impact, it would indicate a crustal temperature above ~1000 K. Most radiogenic systems cannot preserve age information at that temperature owing to isotopic re-setting.[108] For example, the closure temperature for the U-Pb geochronometer in zircon is ~1200 K for a typical 100-200 micron crystal.[108a] The impact energy would likely have reset that clock in the crust.[109] A survey of more than 200,000 detrital zircons from the Jack Hills outcrop in Western Australia shows the oldest at ~4.38 Ga,[51, 107c] which suggests the latest possible date for the veneer-delivering impact.

Mojzsis et al. (2019) place these dates within the context of Solar System-wide events.[71b] Of special importance are estimated dates for the onset of the migration of the giant planets. These estimates relate radiometric ages from asteroidal meteorites to dynamical models that explain late accretion and consider the lack of reset ages after ca. 4.45 Ga.[81a] This may confine the onset of this migration to no later than about 4.48 Ga, with planetesimals and asteroids continuing to strike the inner planets in agreement with crater chronology.

This suggests constraints on $t_0$, the date when the window of opportunity for RNA formation opened, as no earlier than 4.48 Ga, no later than 4.45 Ga, and certainly no later than the oldest reliably dated terran zircons at ~4.36. An impactor of a size able to deliver most of the veneer in one body would have created a magma ocean effectively everywhere on the surface of the Earth.[110] This would have destroyed any complex organic molecules that accumulated earlier. Thus, if 10 million years after a 4.48 Ga impact is needed for a surface to cool enough to allow it to accumulate organics, the window in **Fig. 8** opened at 4.47 Ga. This date would be modestly later if time is required to lose severe greenhouse surface warming.

With $H_2$ loss following a simple exponential with a half-life of 40 million years, atmospheric production of primary nucleobase precursors might be half at 4.44 Ga, a quarter at 4.40 Ga, and an eighth at 4.36 Ga, if the rate of formation of primary precursors is approximated as a linear function of $H_2$ partial pressure. The model may be refined to favor later dates, reflecting (presumably) increasing amounts of sub-aerial land. This provides a non-analytic date for the optimal accumulation of RNA precursors at ~4.36 Ga ± 0.1 Ga. This, according to the model, is the time when the formation of RNA was most probable, as well as the formation of RNA-based Darwinism under an RNA-First model.

### Ways in which these time/date inferences might be wrong

The conclusion that RNA-based life most likely emerged in a sustainable form 4.36 ± 0.1 Ga follows non-analytically from models from physics, geology, and chemistry, as reviewed here. It is also based on a model from biology, that RNA supported the Earth's first Darwinian evolution. However, it is likely to be controversial, in part because it places the origin of life on Earth at a time much earlier than commonly thought. For example, Sutherland, a leader in this field, ties his path-hypotheses and speculations (**Fig 2, Fig. 4**) to the time of the Late Heavy Bombardment at ~3.8 Ga.[111]

Further, the ± 0.1 Ga constraint on the time where RNA formation was most probable is surprisingly narrow. However, its narrowness is an inescapable consequence of physics, once the impact of Moneta is assumed to have set $t_0$. Specifically, it depends on the time constants for the loss of the productivity of the reducing atmosphere as it returns to a redox neutral and unproductive state through $H_2$ loss and mantle outgassing. While these rates have their own uncertainties, these are smaller than the uncertainty in the date of the veneer-delivering impact itself, and the uncertainty of the amount of sub-aerial land. Regardless of those uncertainties, their time constants cannot be hundreds of millions of years, or ten million years or shorter.

Of course, the date inferences, controversial or otherwise, will be incorrect if any of the conclusions from those models from physics, geology, or chemistry are incorrect. We review some of these.

<u>The Hadean mantle may have been more reducing than Trail et al. (2011) inferred.[17]</u>

If the fugacity of the Hadean mantle was substantially below FMQ, then much larger amounts of CO, $H_2$, and $CH_4$ would emerge via outgassing into the atmosphere. This would make the Hadean atmosphere intrinsically more productive, mooting much of the foregoing discussion. In this case, the redox state of the atmosphere fed by a more reduced mantle may be sufficient to support the formation of adequate amounts of RNA precursors, albeit at lower levels than those following a Moneta impact. Balancing this, that productivity might occur over longer periods of time, perhaps increasing the time integrated probability of RNA-based life emerging and being sustained in this way, even though the probability of RNA emerging at any particular time would be lower.

This view is discussed and defended by Yang et al. (2014),[112] and assumed by others to support the prebiotic relevance of experiments run in reducing atmospheres.[42] Measurements of the redox potential of more recent mantles[113] may prove relevant, if we are permitted *a fortiori* extrapolation from the Archean back to the Hadean.

<u>There may be ways to make primary reduced precursors in a redox neutral atmosphere</u>

It is conceivable that our pessimism about the productivity of a redox neutral atmosphere is misplaced. For example, Cleaves et al. (2008)[62] re-assessed prebiotic organic syntheses in neutral planetary atmospheres a decade ago and found more peptide precursors than anticipated under the more pessimistic view. This approach does not appear to have been applied for RNA precursors.

Alternatively, a low productivity of the standard redox-neutral atmosphere can be accepted, if there were another source of energy to drive formation of reduced species. Airapetian et al. (2016)[69] for example suggested that high-energy particles emerging from a very young Sun might be this source. They argue that the early Sun's activity provided a window of opportunity for prebiotic life on Earth. They defined a "biogenic zone" (BZ), within which stellar energy fluxes are high enough to initiate reactive chemistry that produces complex molecules crucial for life. As a by-product, they note that the expected chemistry may form greenhouse gasses that may have kept the atmosphere warmer. This model seems, however, to operate only before the last sterilizing impactor, a sterilization that would have required prebiotic chemistry to "start over" without the benefit of those high-energy particles.

<u>The amount and/or provenance of the Earth and Lunar veneer might be wrong</u>

The conclusions also rely on estimates for the amount of veneer, which (if it was largely delivered in a single impact) sets the size of Moneta at ~$10^{23}$ kg. However, historically, other hypotheses have been proposed to account for the HSE present in Earth's mantle. These include models that postulate inefficient core formation and low metal-silicate partition ratios at elevated temperatures and pressures at the base of a deep magma ocean. These have been reviewed with their various strengths and weaknesses using the large database now available covering measurements of siderophiles in the terrestrial mantle, lunar samples, and martian meteorites.[79] While other factors may have contributed, the view that most of the veneer arrived from late impacts seems secure.

The conclusion that RNA-based life most likely emerged in a sustainable form 4.36 ± 0.1 Ga also relies on the statistics of small numbers to account for the surprisingly little veneer on the Lunar surface. This drives the model having a single large impactor with a mass of $10^{23}$ kg deliver the veneer, rather than many small impactors.

However, other models may explain the low veneer on the Moon relative to Earth.[83c] For example, Sleep (2016)[114] accounts for the differences in Terran and Lunar veneers with a model that has the Lunar veneer moving into the Lunar iron core without an analogous process on Earth. In this model, the veneer arises with the Moon-forming impactor, offering no impact-related opportunity for RNA to form at that time, as the Moon-forming impact more-than-destroyed any organics on the Earth's surface.

Alternatively, Kraus et al. (2015)[83a] suggest that the modest Lunar veneer is a consequence of selective escape of iron (presumably with siderophiles) from the Moon due to its lower escape velocity. If this model is correct, iron would still have been delivered to the Hadean atmosphere in smaller impactors, lowering its redox potential. However, that lowering would have been less dramatic, the productivity would not have been as high, and integrated probability for RNA formation would be lower.

Of course, the measured estimates of the amount of the Lunar veneer might be simply incorrect. The amount of the veneer in the Lunar mantle must be inferred from analyses of derived melts,[79] meaning that fractionation in that derivation must be inferred. For example, if the melt is sufficiently oxidized, rhenium and iridium might be lost upon volcanic eruption.[115] The likelihood of this view decreases as our inventory of samples grows.

<u>An impactor *after* Moneta may have been sterilizing</u>

This statement about probability is based on a particular model for the impact history of Earth. Specifically, that model views that an impact that delivered essentially all of the late veneer was also the *last* impact that was planet-sterilizing. A later, smaller impactor might have sterilized the Earth without making a noticeable contribution to the late veneer, if it was ~$10^{21}$ kg,[116] the smallest impactor likely to have been globally sterilizing. An impactor this size would deliver only ~1% of the late veneer. Thus, and sequentially, the impact history may (or may not) involve impacts that re-form the core (Theia, for example), those too small to re-form the core but large enough to re-set the mantle radiological clocks (Moneta, for example), those too small to re-set the clocks but large enough to be globally sterilizing (but leave no record in the veneer), those that made a globally productive atmosphere but were not globally sterilizing, and last, the smallest, those that created a locally productive environment. A set of possible impactors is collected in **Table 4**.

**Table 4**. Impacts and impactors relevant to the discussion*

| Impactor name | Size similar to | Mass (kg) | Sterilizing? | When | Geological evidence |
|---|---|---|---|---|---|
| Theia | modern Mars | $8.3 \times 10^{24}$ | very | ~4.51 Ga | U-Pb dates from the Moon, *inter alia*[49d] |
| Moneta | modern Moon | $7.4 \times 10^{22}$ | yes | ~4.47 Ga | Last global resetting of clocks; major contribution to the late veneer. |
| Unnamed | modern Ceres | $9.0 \times 10^{20}$ | Possibly† | Most likely before ~4.35 Ga | Too small to have left any significant reset ages. Negligible contribution to the veneer. |
| Unnamed | modern Vesta | $2.4 \times 10^{20}$ | Only locally | Most likely before ~4.29 Ga | Much too small to have left any significant reset ages. Very negligible contribution to the veneer. |
| Late heavy bombardment | unknown | unknown | unknown | Probably never | Re-interpretation of lunar crater record, meteorite reset ages, inter alia[71b, 117] |

* Zahnle and Carlson have an analogous table in preparation (Kevin Zahnle, personal communication).
† The degree to which life was able to survive an impact in refugia (for example, deep ocean or subsurface) depends in part on the rate at which life evolved to inhabit such locales; that rate is very much unknown for RNA-based life, notwithstanding the evident adaptability of the protein-RNA life of the modern world.

If Moneta were not the last globally sterilizing impactor, then RNA life would have emerged most probably in the (shorter) window of opportunity created by the (smaller) last sterilizing impactor. This would alter our discussion of probabilities and timelines in two ways, (i) by moving $t_0$ to a later time, and (ii) by requiring that a smaller inflow of reduced precursors be adequate to support the prebiotic synthesis of RNA and, subsequently, sustain RNA-based Darwinian evolution, at least long enough for it to evolve metabolic pathways to replace food it had previously obtained by precipitation.

Neither alteration is major. Assuming a succession of impactors of decreasing flux,[118] later impactors would, on average, have opened smaller windows of opportunity with less productive atmospheres, due to their smaller iron cores. The windows of prebiotic productivity would have been shorter, opening sooner after $t_0$ but disappearing even sooner. Also diminishing would be the ability of the atmosphere to continue to feed the nascent RNA organisms; they would need to have evolved their metabolism more rapidly.

We can estimate $t_0$'s here from models of the size of impacts that would be sterilizing without leaving refugia. To be globally sterilizing, an impact must melt the crust essentially all at once.[119] The associated melt front from such an impact must extend at least 4 km into the crust, assuming that whatever life that was present had evolved to occupy those habitats. This requires adequate energy to locally evaporate the oceans, and then more to create a melt front. The analyses of Abramov & Mojzsis (2009)[116, 120] suggest that such an object must be somewhat larger than Vesta ($2.6 \times 10^{20}$ kg) just to evaporate the oceans. To comprehensively melt the crust, a larger Ceres-sized impactor is necessary (~$10^{21}$ kg).

How many $10^{21}$ kg impacts likely occurred after the Moneta impact? Models, as well as the Lunar and Martian cratering records,[71a, 118] the expected mass of the asteroid belt, and the siderophile abundances in the Moon,[71a] all suggest that the number of Vesta-sized impacts onto the Earth was 1 ± 1.[71a, 82] That is, a later sterilizing impactor alternative occurred maybe just once, maybe twice, or maybe not at all. The most probable time for the "once" scenario was ~4443 Ma, with a 95% chance that it happened before 4291 Ma. A non-sterilizing impactor of mass $3 \times 10^{19}$ kg would create ~1 bar of $H_2$; models expect roughly 1.5 such events over the course of Earth's history.

In addition to having not left a signature in the veneer, such smaller impacts (if they occurred at all) evidently left no evidence in the zircon record. The existence of material with reliable radiogenic dates older than 4.35 Ga implies that no impactor-resets occurred after that date. Of course, a $10^{22}$ kg impact would not, even though it would have had sterilizing potential, and also have created a productive atmosphere. The Pb-Pb ages of the Earth are all 4480 Ma.[121] This age is consistent with the Moneta-based natural history outlined here; its size would have likely re-set (most) clocks; the next smaller impactor in **Table 4** would not.

In the last category, it is possible for even a very small impactor (diameter ~50 km) to *locally* generate conditions productive for RNA precursor formation. For example, Parkos *et al*. (2018) suggested that impact ejecta may have produced sufficiently concentrated HCN in shallow bodies of water to support prebiotic synthesis, especially if the atmosphere was already partially reducing.[89]e However, the time integrated probability of RNA emerging from one of these very small impacts on an uncertain, possibly small, amount of dry land, seems smaller than the time integrated probability of RNA emerging from an event that is planet-wide, and therefore delivers RNA precursors to *any* dry land that exists.

## Summary and Outlook

The widespread presence of RNA cofactors, the ribosome, and catalytic RNA on modern Earth and in the last common ancestor (and likely before)[1e] makes an RNA-First model for the origin of life worth considering (at least). Most chemical path-hypotheses to create RNA prebiologically all but require an atmosphere that is more reducing than what planetary accretion models suggest was the norm above the Hadean Earth. Even serpentinization, if

e Impacts in a "late heavy bombardment" (LHB) were thought to lie outside of a uniformly decreasing size and/or frequency distribution; the LHB was thought to be an exceptional burst of large, possibly sterilizing impacts at ~3.9 Ga. The initial argument for a LHB, based in part on dates of non-random samples returned by Apollo, has lost its following given re-analyses of Moon-derived meteorites, impact records of other Solar System bodies, and a reassessment of the Lunar record. Current views hold that the LHB, if it existed at all, was not planet-sterilizing.[117, 122] This requires reconsideration of the dates of models based on it.[18, 22] it creates only tens of parts per million of methane to the atmosphere,[75] would not have been sufficient.

This, in turn, requires a mechanism to add reducing power to a redox-neutral Hadean atmosphere that, if dominated by outgassing of the most likely mantle, would not have delivered the needed reduced nitrogen-containing RNA precursors. While alternative mechanisms exist for introducing some reducing power into the atmosphere, a $10^{23}$ kg impactor (Moneta) would do so in large amounts, and for a long time. Although the inference is not analytic, this impact scenario appears to be the most probable way of creating the reduced precursors needed to generate and sustain RNA-based life.

This impact, independently, accounts for many facts from geology. These include the amounts of siderophilic elements in the veneer on the Earth and Moon, measurements of many kinds on zircons surviving from the Hadean, and the Lunar (and Martian) cratering record. Further, it is consistent with our evolving view of the physics of the Solar System, including how that physics translates into a model for the natural history of impacts throughout. This, in turn, is consistent with data showing a general resetting of radioisotopic clocks on Earth at 4.48 Ga, which would not be done by an impactor an order of magnitude smaller than $10^{23}$ kg.

If this $10^{23}$ kg Moneta impact sets $t_0$ at 4.48 Ga, considerations of the rate of congealing of magma oceans, dissipation of various greenhouse effects to bring the surface to a reasonable temperature, and the need to accumulate reservoirs of RNA precursors, all suggest that the window of opportunity for the prebiotic formation of RNA opened a short time afterwards. Assuming a maximum at three half-lives (each 40 Myr) for the subsequent decline in productivity of the atmosphere, this suggests that the most probable date for RNA to have been formed is ca. 4.36 ± 0.1 Ga. The error bar on that number (± 100 Myr) is not analytic, but it cannot be much larger; restoration of the redox neutral atmosphere did not take hundreds of millions of years, and RNA would not have survived on a hot planet very soon after the Moneta impact. Indeed, the principal source of uncertainty is the amount of subaerial land that was available to collect the reservoirs of precipitating RNA precursors.

It is worth noting that the date that we conjecture is some 250 million years *before* the 4.1 Ga date of a single zircon that has been reported to contain carbon inclusions depleted in $^{13}$C.[53] This depletion is reminiscent of the depletion seen in modern carbon having a biological origin. Again, as reviewed above, the origin of the $^{13}$C depleted carbon is highly controversial, and both we and the reporting observers discuss these data with a full range of caveats. Further, it does not drive the discussion above. However, *if* that $^{13}$C level reflects a biological source, it is interesting to note that Earth would have had ca. 250 million years after RNA-based Darwinian evolution most probably arose to have accumulated a detectable biosphere by 4.1 Ga.

## Acknowledgements

We are indebted to David Fialho (Georgia Institute of Technology, Atlanta GA), Paul Higgs (McMaster University, Hamilton Ontario), Ram Krishnamurthy (The Scripps Research Institute, La Jolla CA), Matthew Powner (University College London), Sukrit Ranjan (MIT, Cambridge MA), Zoe R. Todd (Harvard-Smithsonian Center for Astrophysics, Cambridge MA), and Loren D. Williams (Georgia Institute of Technology, Atlanta GA), all of whom participated in the discussion in Atlanta on October 17, 2018, and contributed ideas and comments for this review. We are indebted to Lucy Kwok of the Earth Life Science Institute at Tokyo Institute of Technology for preparing Figure 8. We are also indebted to David Catling, James Kasting, and Kevin Zahnle for extensive criticisms of the manuscript. This publication was made possible through the support of a grant from the John Templeton Foundation (54466). The opinions expressed in this publication are those of the authors and do not necessarily reflect the views of the John Templeton Foundation.

**Keywords:** geochemistry • origin of life • oxidation • prebiotic chemistry • RNA world

### References

[1] a) H. B. White, *J. Mol. Evol.* **1976**, *7*, 101-104; b) C. M. Visser, R. M. Kellogg, *J. Mol. Evol.* **1978**, *11*, 163-168; c) C. M. Visser, R. M. Kellogg, *J. Mol. Evol.* **1978**, *11*, 171-187; d) W. Gilbert, *Nature* **1986**, *319*, 618-618; e) S. A. Benner, A. D. Ellington, A. Tauer, *Proc. Natl. Acad. Sci. U.S.A.* **1989**, *86*, 7054-7058.

[2] M. Neveu, H. J. Kim, S. A. Benner, *Astrobiology* **2013**, *13*, 391-403.

[3] C. Briones, M. Stich, S. C. Manrubia, *RNA* **2009**, *15*, 743-749.

[4] a) M. W. Powner, B. Gerland, J. D. Sutherland, *Nature* **2009**, *459*, 239-242; b) H. J. Kim, A. Ricardo, H. I. Illangkoon, M. J. Kim, M. A. Carrigan, F. Frye, S. A. Benner, *J. Am. Chem. Soc.* **2011**, *133*, 9457-9468; c) S. A. Benner, H. J. Kim, M. A. Carrigan, *Acc. Chem. Res.* **2012**, *45*, 2025-2034; d) S. Becker, I. Thoma, A. Deutsch, T. Gehrke, P. Mayer, H. Zipse, T. Carell, *Science* **2016**, *352*, 833-836.

[5] N. Kitadai, S. Maruyama, *Geosci. Front.* **2018**, *9*, 1117-1153.

[6] a) N. V. Hud, B. J. Cafferty, R. Krishnamurthy, L. D. Williams, *Chem. Biol.* **2013**, *20*, 466-474; b) K. A. Lanier, A. S. Petrov, L. D. Williams, *J. Mol. Evol.* **2017**, *85*, 8-13; c) C. M. Runnels, K. A. Lanier, J. K. Williams, J. C. Bowman, A. S. Petrov, N. V. Hud, L. D. Williams, *J. Mol. Evol.* **2018**, *86*, 598-610; d) D. M. Fialho, K. C. Clarke, M. K. Moore, G. B. Schuster, R. Krishnamurthy, N. V. Hud, *Org. Biomol. Chem.* **2018**, *16*, 1263-1271.

[7] C. Menor-Salvan, D. M. Ruiz-Bermejo, M. I. Guzman, S. Osuna-Esteban, S. Veintemillas-Verdaguer, *Chemistry* **2009**, *15*, 4411-4418.

[8] A. Pross, *Orig. Life Evol. Biosph.* **2004**, *34*, 307-321.

[9] J. Kawai, D. C. McLendon, H. J. Kim, S. A. Benner, *Astrobiology* **2019**, *19*, 506-516.

[10] R. Krishnamurthy, S. Guntha, A. Eschenmoser, *Angew. Chem. Int. Ed. Engl.* **2000**, *39*, 2281-2285.

[11] a) H. J. Kim, S. A. Benner, *Proc. Natl. Acad. Sci. U.S.A.* **2017**, *114*, 11315-11320; b) H. J. Kim, J. Kim, *Astrobiology* **2019**, *19*, 669-674.

[12] N. G. Holm, M. Dumont, M. Ivarsson, C. Konn, *Geochem. Trans.* **2006**, *7*, 7.

[13] a) J. A. da Silva, N. G. Holm, *J. Colloid. Interface Sci.* **2014**, *431*, 250-254; b) Y. Furukawa, H. J. Kim, D. Hutter, S. A. Benner, *Astrobiology* **2015**, *15*, 259-267; c) H. J. Kim, Y. Furukawa, T. Kakegawa, A. Bita, R. Scorei, S. A. Benner, *Angew. Chem. Int. Ed. Engl.* **2016**, *55*, 15816-15820.

[14] a) C. Reid, L. E. Orgel, *Nature* **1967**, *216*, 455; b) I. Parsons, M. R. Lee, J. V. Smith, *Proc. Natl. Acad. Sci. U.S.A.* **1998**, *95*, 15173-15176; c) J. V. Smith, F. P. Arnold, Jr., I. Parsons, M. R. Lee, *Proc. Natl. Acad. Sci. U.S.A.* **1999**, *96*, 3479-3485; d) F. Westall, K. Hickman-Lewis, N. Hinman, P. Gautret, K. A. Campbell, J. G. Breheret, F. Foucher, A. Hubert, S. Sorieul, A. V. Dass, T. P. Kee, T. Georgelin, A. Brack, *Astrobiology* **2018**, *18*, 259-293.

[15] J. L. Bada, J. Korenaga, *Life (Basel)* **2018**, *8*.

[16] L. E. Orgel, *Crit. Rev. Biochem. Mol. Biol.* **2004**, *39*, 99-123.

[17] D. Trail, E. B. Watson, N. D. Tailby, *Nature* **2011**, *480*, 79-82.

[18] D. J. Ritson, C. Battilocchio, S. V. Ley, J. D. Sutherland, *Nat. Commun.* **2018**, *9*, 1821.

[19] H. Okamura, S. Becker, N. Tiede, S. Wiedemann, J. Feldmann, T. Carell, *Chem. Commun. (Camb.)* **2019**, *55*, 1939-1942.

[20] S. Ranjan, Z. R. Todd, P. B. Rimmer, D. D. Sasselov, A. R. Babbin, *Geochem. Geophy. Geosy.* **2019**, *20*, 2021-2039.

[21] A. D. Keefe, S. L. Miller, *Origins of Life and Evolution of the Biosphere* **1996**, *26*, 111-129.

[22] J. Xu, D. J. Ritson, S. Ranjan, Z. R. Todd, D. D. Sasselov, J. D. Sutherland, *Chem. Commun. (Camb.)* **2018**, *54*, 5566-5569.

[23] a) R. Lohrmann, L. E. Orgel, *Science* **1971**, *171*, 490-494; b) M. J. Bishop, R. Lohrmann, L. E. Orgel, *Nature* **1972**, *237*, 162-164; c) S. A. Benner, *Nat. Commun.* **2018**, *9*, 5173.

[24] H. Slebocka-Tilk, F. Sauriol, M. Monette, R. S. Brown, *Can. J. Chem.* **2002**, *80*, 1343-1350.

[25] J. D. Bernal, *The Physical Basis of Life*, Routledge and Kegan Paul, London, **1951**.
[26] J. P. Ferris, A. R. Hill, Jr., R. Liu, L. E. Orgel, *Nature* **1996**, *381*, 59-61.
[27] H. G. Hansma, *Orig. Life Evol. Biosph.* **2014**, *44*, 307-311.
[28] a) J. Oro, *Biochem. Bioph. Res. Co.* **1960**, *2*, 407-412; b) J. Oro, A. P. Kimball, *Arch. Biochem. Biophys.* **1961**, *94*, 217-227.
[29] G. Steinman, R. M. Lemmon, M. Calvin, *Proc. Natl. Acad. Sci. U.S.A.* **1964**, *52*, 27-30.
[30] a) R. A. Sanchez, J. P. Ferris, L. E. Orgel, *Science* **1966**, *154*, 784-785; b) M. P. Robertson, S. L. Miller, *Nature* **1995**, *375*, 772-774.
[31] Y. L. Wang, H. D. Lee, M. W. Beach, D. W. Margerum, *Inorg. Chem.* **1987**, *26*, 2444-2449.
[32] a) J. P. Ferris, E. A. Williams, D. E. Nicodem, J. S. Hubbard, G. E. Voecks, *Nature* **1974**, *249*, 437-439; b) R. Hayatsu, M. H. Studier, A. Oda, K. Fuse, E. Anders, *Geochim. Cosmochim. Ac.* **1968**, *32*, 175-&.
[33] B. Burcar, M. Pasek, M. Gull, B. J. Cafferty, F. Velasco, N. V. Hud, C. Menor-Salvan, *Angew. Chem. Int. Ed. Engl.* **2016**, *55*, 13249-13253.
[34] a) R. Saladino, M. Barontini, C. Cossetti, E. Di Mauro, C. Crestini, *Orig. Life Evol. Biosph.* **2011**, *41*, 317-330; b) R. Saladino, U. Ciambecchini, C. Crestini, G. Costanzo, R. Negri, E. Di Mauro, *Chembiochem.* **2003**, *4*, 514-521.
[35] A. M. Schoffstall, R. J. Barto, D. L. Ramos, *Orig. Life* **1982**, *12*, 143-151.
[36] B. Gedulin, Arrhenius, G., in *Early life on Earth – Nobel symposium 84* (Ed.: S. Bengston), Columbia University Press, New York, **1994**, pp. 91-110.
[37] A. Ricardo, M. A. Carrigan, A. N. Olcott, S. A. Benner, *Science* **2004**, *303*, 196.
[38] E. Biondi, L. Howell, S. A. Benner, *Synlett.* **2018**, *29*, 256-256.
[39] E. Biondi, Y. Furukawa, J. Kawai, S. A. Benner, *Beilstein J. Org. Chem.* **2017**, *13*, 393-404.
[40] E. W. Ziegler, H. J. Kim, S. A. Benner, *Astrobiology* **2018**, *18*, 1159-1170.
[41] M. A. Pasek, *Geosci. Front.* **2017**, *8*, 329-335.
[42] M. Ferus, F. Pietrucci, A. M. Saitta, A. Knizek, P. Kubelik, O. Ivanek, V. Shestivska, S. Civis, *Proc. Natl. Acad. Sci. U.S.A.* **2017**, *114*, 4306-4311.
[43] a) A. A. Borisov, *Petrology+* **2016**, *24*, 117-126; b) F. Farges, R. Siewert, G. E. Brown, A. Guesdon, G. Morin, *Can. Mineral.* **2006**, *44*, 731-753; c) F. Farges, R. Siewert, C. W. Ponader, G. E. Brown, M. Pichavant, H. Behrens, *Can. Mineral.* **2006**, *44*, 755-773.
[44] B. Herschy, S. J. Chang, R. Blake, A. Lepland, H. Abbott-Lyon, J. Sampson, Z. Atlas, T. P. Kee, M. A. Pasek, *Nat. Commun.* **2018**, *9*, 1346.
[45] a) N. Dauphas, *Nature* **2017**, *541*, 521-524; b) R. Brasser, N. Dauphas, S. J. Mojzsis, *Geophys. Res. Lett.* **2018**, *45*, 5908-5917.
[46] K. Lodders, *Space Sci. Rev.* **2000**, *92*, 341-354.
[47] D. J. Stevenson, *Science* **1981**, *214*, 611-619.
[48] a) B. J. Wood, A. N. Halliday, *Nature* **2010**, *465*, 767-770; b) T. Kleine, J. F. Rudge, *Elements* **2011**, *7*, 41-46.
[49] a) C. J. Allegre, G. Manhes, C. Gopel, *Geochim. Cosmochim. Ac.* **1995**, *59*, 1445-1456; b) M. D. Norman, L. E. Borg, L. E. Nyquist, D. D. Bogard, *Meteorit. Planet. Sci.* **2003**, *38*, 645-661; c) S. Charnoz, C. Michaut, *Icarus* **2015**, *260*, 440-463; d) M. Barboni, P. Boehnke, B. Keller, I. E. Kohl, B. Schoene, E. D. Young, K. D. McKeegan, *Sci. Adv.* **2017**, *3*, e1602365.
[50] J. Wade, B. J. Wood, *Earth Planet. Sc. Lett.* **2005**, *236*, 78-95.
[51] T. M. Harrison, E. A. Bell, P. Boehnke, *Petrochronology: Methods and Applications* **2017**, *83*, 329-+.
[52] a) S. J. Mojzsis, T. M. Harrison, R. T. Pidgeon, *Nature* **2001**, *409*, 178-181; b) A. J. Cavosie, J. W. Valley, S. A. Wilde, *Earth Planet. Sc. Lett.* **2005**, *235*, 663-681; c) T. T. Isson, N. J. Planavsky, *Nature* **2018**, *560*, 471-475.
[53] E. A. Bell, P. Boehnke, T. M. Harrison, W. L. Mao, *Proc. Natl. Acad. Sci. U.S.A.* **2015**, *112*, 14518-14521.
[54] a) B. Rasmussen, I. R. Fletcher, J. R. Muhling, C. J. Gregory, S. A. Wilde, *Geology* **2011**, *39*, 1143-1146; b) M. Hopkins, T. M. Harrison, C. E. Manning, *Geology* **2012**, *40*, E281-E281; c) J. D. Vervoort, A. I. S. Kemp, *Chem. Geol.* **2016**, *425*, 65-75; d) M. J. Whitehouse, A. A. Nemchin, R. T. Pidgeon, *Gondwana Res.* **2017**, *51*, 78-91; e) J. Alleon, R. E. Summons, *Free Radic. Biol. Med.* **2019**.
[55] S. Moorbath, *Nature* **2005**, *434*, 155.
[56] S. J. Mojzsis, G. Arrhenius, K. D. McKeegan, T. M. Harrison, A. P. Nutman, C. R. Friend, *Nature* **1996**, *384*, 55-59.
[57] Y. Ohtomo, T. Kakegawa, A. Ishida, T. Nagase, M. T. Rosing, *Nat. Geosci.* **2014**, *7*, 25-28.
[58] B. Sherwood Lollar, T. D. Westgate, J. A. Ward, G. F. Slater, G. Lacrampe-Couloume, *Nature* **2002**, *416*, 522-524.
[59] K. Hamano, Y. Abe, H. Genda, *Nature* **2013**, *497*, 607-610.
[60] J. F. Kasting, *Science* **1993**, *259*, 920-926.
[61] M. M. Hirschmann, *Earth Planet. Sc. Lett.* **2012**, *341*, 48-57.
[62] H. J. Cleaves, J. H. Chalmers, A. Lazcano, S. L. Miller, J. L. Bada, *Orig. Life Evol. Biosph.* **2008**, *38*, 105-115.
[63] a) C. E. Harman, J. F. Kasting, E. T. Wolf, *Orig. Life Evol. Biosph.* **2013**, *43*, 77-98; b) K. J. Zahnle, *Elements* **2006**, *2*, 217-222.
[64] J. P. Pinto, G. R. Gladstone, Y. L. Yung, *Science* **1980**, *210*, 183-185.
[65] D. C. Rubie, D. J. Frost, U. Mann, Y. Asahara, F. Nimmo, K. Tsuno, P. Kegler, A. Holzheid, H. Palme, *Earth Planet. Sc. Lett.* **2011**, *301*, 31-42.
[66] a) K. J. Zahnle, *J. Geophys. Res.-Atmos.* **1986**, *91*, 2819-2834; b) F. Tian, J. F. Kasting, K. Zahnle, *Earth Planet. Sc. Lett.* **2011**, *308*, 417-423.
[67] a) C. N. Matthews, R. D. Minard, *Faraday Discuss.* **2006**, *133*, 393-401; discussion 427-352; b) B. K. D. Pearce, R. E. Pudritz, D. A. Semenov, T. K. Henning, *Proc. Natl. Acad. Sci. U.S.A.* **2017**, *114*, 11327-11332.
[68] K. Kurosawa, S. Sugita, K. Ishibashi, S. Hasegawa, Y. Sekine, N. O. Ogawa, T. Kadono, S. Ohno, N. Ohkouchi, Y. Nagaoka, T. Matsui, *Orig. Life Evol. Biosph.* **2013**, *43*, 221-245.
[69] V. S. Airapetian, A. Glocer, G. Gronoff, E. Hebrard, W. Danchi, *Nat. Geosci.* **2016**, *9*, 452-+.
[70] C. Chyba, C. Sagan, *Nature* **1992**, *355*, 125-132.
[71] a) R. Brasser, S. C. Werner, S. J. Mojzsis, *Meteorit. Planet. Sci.* **2019**, *54*; b) S. J. Mojzsis, R. Brasser, N. M. Kelly, O. Abramov, S. C. Werner, *Astrophys. J.* **2019**, *881*, 44-56.
[72] J. R. Lyons, C. Manning, F. Nimmo, *Geophys. Res. Lett.* **2005**, *32*.
[73] J. F. Kasting, D. Catling, *Annu. Rev. Astron. Astr.* **2003**, *41*, 429-463.
[74] a) W. L. Chameides, J. C. Walker, *Orig. Life* **1981**, *11*, 291-302; b) S. L. Miller, G. Schlesinger, *Origins of Life and Evolution of the Biosphere* **1984**, *14*, 83-90.
[75] J. F. Kasting, *Geol. Soc. Am. Spec.* **2014**, *504*, 19-28.
[76] S. L. Miller, G. Schlesinger, *Adv. Space Res.* **1983**, *3*, 47-53.
[77] D. J. Ritson, J. D. Sutherland, *Angew. Chem. Int. Ed. Engl.* **2013**, *52*, 5845-5847.
[78] J. M. D. Day, A. D. Brandon, R. J. Walker, *Rev. Mineral Geochem.* **2016**, *81*, 161-238.
[79] R. J. Walker, *Chem. Erde-Geochem.* **2009**, *69*, 101-125.
[80] a) M. Willbold, S. J. Mojzsis, H. W. Chen, T. Elliott, *Earth Planet. Sc. Lett.* **2015**, *419*, 168-177; b) J. M. Day, D. G. Pearson, L. A. Taylor, *Science* **2007**, *315*, 217-219.
[81] a) W. F. Bottke, D. Vokrouhlicky, S. Marchi, T. Swindle, E. R. Scott, J. R. Weirich, H. Levison, *Science* **2015**, *348*, 321-323; b) W. F. Bottke, R. J. Walker, J. M. Day, D. Nesvorny, L. Elkins-Tanton, *Science* **2010**, *330*, 1527-1530; c) D. C. Rubie, V. Laurenz, S. A. Jacobson, A. Morbidelli, H. Palme, A. K. Vogel, D. J. Frost, *Science* **2016**, *353*, 1141-1144; d) E. A. Frank, W. D. Maier, S. J. Mojzsis, *Contrib. Mineral. Petr.* **2016**, *171*; e) H. Genda, R. Brasser, S. J. Mojzsis, *Earth Planet. Sc. Lett.* **2017**, *480*, 25-32.
[82] R. Brasser, S. J. Mojzsis, S. C. Werner, S. Matsumura, S. Ida, *Earth Planet. Sc. Lett.* **2016**, *455*, 85-93.
[83] a) R. G. Kraus, S. Root, R. W. Lemke, S. T. Stewart, S. B. Jacobsen, T. R. Mattsson, *Nat. Geosci.* **2015**, *8*, 269-272; b) H. E. Schlichting, E. O. Ofek, R. Sari, E. P. Nelan, A. Gal-Yam, M. Wenz, P. Muirhead, N. Javanfar, M. Livio, *Astrophys. J.* **2012**, *761*; c) H. E. Schlichting, P. H. Warren, Q. Z. Yin, *Astrophys. J.* **2012**, *752*.
[84] J. M. D. Day, R. J. Walker, *Earth Planet. Sc. Lett.* **2015**, *423*, 114-124.
[85] a) R. M. Canup, E. Asphaug, *Nature* **2001**, *412*, 708-712; b) H. J. Melosh, *Impact Cratering: A Geologic Process.*, Oxford University Press, New York, **1989**.
[86] C. J. Allegre, G. Manhes, C. Gopel, *Earth Planet. Sc. Lett.* **2008**, *267*, 386-398.
[87] H. Genda, T. Iizuka, T. Sasaki, Y. Ueno, M. Ikoma, *Earth Planet. Sc. Lett.* **2017**, *470*, 87-95.
[88] a) T. Staudacher, C. J. Allegre, *Earth Planet. Sc. Lett.* **1982**, *60*, 389-406; b) F. Albarede, J. Martine, *Geochim. Cosmochim. Ac.* **1984**, *48*, 207-212.
[89] D. Parkos, A. Pikus, A. Alexeenko, H. J. Melosh, *J. Geophys. Res.-Planet.* **2018**, *123*, 892-909.
[90] a) G. Toupance, F. Raulin, R. Buvet, *Orig. Life* **1975**, *6*, 83-90; b) F. Raulin, G. Toupance, in *Cosmochemical Evolution and the Origins of Life*, Springer, Dordrecht, **1974**, pp. 91-97.
[91] R. Stribling, S. L. Miller, *Orig. Life Evol. Biosph.* **1987**, *17*, 261-273.
[92] H. S. C. O'Neill, H. Palme, in *The Earth's Mantle; Composition, Structure, and Evolution.* (Ed.: I. Jackson), Cambridge University Press, Cambridge, **1998**, pp. 3-126.
[93] M. A. Pasek, T. P. Kee, D. E. Bryant, A. A. Pavlov, J. I. Lunine, *Angew. Chem. Int. Ed. Engl.* **2008**, *47*, 7918-7920.
[94] S. Miyakawa, H. J. Cleaves, S. L. Miller, *Orig. Life Evol. Biosph.* **2002**, *32*, 195-208.
[95] R. P. Welcher, M. E. Castellion, V. P. Wystrach, *J. Am. Chem. Soc.* **1959**, *81*, 2541-2547.
[96] a) M. J. Russell, A. J. Hall, *J. Geol. Soc. London* **1997**, *154*, 377-402; b) M. J. Russell, A. J. Hall, W. Martin, *Geobiology* **2010**, *8*, 355-371.
[97] a) R. Wordsworth, R. Pierrehumbert, *Science* **2013**, *339*, 64-67; b) R. Wordsworth, Y. Kalugina, S. Lokshtanov, A. Vigasin, B. Ehlmann, J. Head, C. Sanders, H. Wang, *Geophys. Res. Lett.* **2017**, *44*, 665-671.
[98] a) M. Hopkins, T. M. Harrison, C. E. Manning, *Nature* **2008**, *456*, 493-496; b) M. D. Hopkins, T. M. Harrison, C. E. Manning, *Earth Planet. Sc. Lett.* **2010**, *298*, 367-376; c) E. A. Bell, P. Boehnke, T. Mark Harrison, *Earth Planet. Sc. Lett.* **2017**, *473*, 237-246.
[99] J. Korenaga, *Philos. Trans. A Math. Phys. Eng. Sci.* **2018**, *376*.
[100] G. Caro, P. Morino, S. J. Mojzsis, N. L. Cates, W. Bleeker, *Earth Planet. Sc. Lett.* **2017**, *457*, 23-37.
[101] J. Korenaga, N. J. Planavsky, D. A. D. Evans, *Philos. Trans. A Math. Phys. Eng. Sci.* **2017**, *375*.
[102] I. Ribas, G. F. P. de Mello, L. D. Ferreira, E. Hebrard, F. Selsis, S. Catalan, A. Garces, J. D. do Nascimento, J. R. de Medeiros, *Astrophys. J.* **2010**, *714*, 384-395.
[103] a) G. Arney, S. D. Domagal-Goldman, V. S. Meadows, E. T. Wolf, E. Schwieterman, B. Charnay, M. Claire, E. Hebrard, M. G. Trainer,

*Astrobiology* **2016**, *16*, 873-899; b) G. N. Arney, *Astrophys. J. Lett.* **2019**, *873*; c) G. N. Arney, V. S. Meadows, S. D. Domagal-Goldman, D. Deming, T. D. Robinson, G. Tovar, E. T. Wolf, E. Schwieterman, *Astrophys. J.* **2017**, *836*.
[104] R. Shapiro, *Proc. Natl. Acad. Sci. U.S.A.* **1999**, *96*, 4396-4401.
[105] N. Guttenberg, N. Virgo, K. Chandru, C. Scharf, I. Mamajanov, *Philos. Trans. A Math. Phys. Eng. Sci.* **2017**, *375*.
[106] W. R. Kuhn, S. K. Atreya, *Icarus* **1979**, *37*, 207-213.
[107] a) A. S. G. Roth, B. Bourdon, S. J. Mojzsis, J. F. Rudge, M. Guitreau, J. Blichert-Toft, *Geochem. Geophy. Geosy.* **2014**, *15*, 2329-2345; b) T. M. Harrison, *Annu. Rev. Earth Pl. Sc.* **2009**, *37*, 479-505; c) J. W. Valley, A. J. Cavosie, T. Ushikubo, D. A. Reinhard, D. F. Lawrence, D. J. Larson, P. H. Clifton, T. F. Kelly, S. A. Wilde, D. E. Moser, M. J. Spicuzza, *Nat. Geosci.* **2014**, *7*, 219-223; d) L. C. Bouvier, M. M. Costa, J. N. Connelly, N. K. Jensen, D. Wielandt, M. Storey, A. A. Nemchin, M. J. Whitehouse, J. F. Snape, J. J. Bellucci, F. Moynier, A. Agranier, B. Gueguen, M. Schonbachler, M. Bizzarro, *Nature* **2018**, *558*, 586-589.
[108] a) D. J. Cherniak, *Rev. Mineral. Geochem.* **2010**, *72*, 827-869; b) D. J. Cherniak, *Rev. Mineral. Geochem.* **2010**, *72*, 691-733; c) D. J. Cherniak, A. Dimanov, *Rev. Mineral. Geochem.* **2010**, *72*, 641-690.
[109] O. Abramov, D. A. Kring, S. J. Mojzsis, *Chem. Erde-Geochem.* **2013**, *73*, 227-248.
[110] N. H. Sleep, K. Zahnle, *J. Geophys. Res.-Planet.* **2001**, *106*, 1373-1399.
[111] J. D. Sutherland, *Angew. Chem. Int. Ed. Engl.* **2016**, *55*, 104-121.
[112] X. Z. Yang, F. Gaillard, B. Scaillet, *Earth Planet. Sc. Lett.* **2014**, *393*, 210-219.
[113] a) Y. Moussallam, G. Oppenheimer, B. Scaillet, *Earth Planet. Sc. Lett.* **2019**, *520*, 260-267; b) R. W. Nicklas, I. S. Puchte, R. D. Ash, H. M. Piccoli, E. Hanski, E. G. Nisbet, P. Waterton, D. G. Pearson, A. D. Anbar, *Geochim. Cosmochim. Ac.* **2019**, *250*, 49-75; c) S. Aulbach, V. Stagno, *Geology* **2016**, *44*, 751-754.
[114] N. H. Sleep, *Geochem. Geophy. Geosy.* **2016**, *17*, 2623-2642.
[115] a) W. Sun, V. C. Bennett, S. M. Eggins, V. S. Kamenetsky, R. J. Arculus, *Nature* **2003**, *422*, 294-297; b) J. C. Lassiter, J. Blichert-Toft, E. H. Hauri, H. G. Barsczus, *Chem. Geol.* **2003**, *202*, 115-138.
[116] O. Abramov, S. J. Mojzsis, *Geochim. Cosmochim. Ac.* **2009**, *73*, A5-A5.
[117] N. E. B. Zellner, *Orig. Life Evol. Biosph.* **2017**, *47*, 261-280.
[118] S. C. Werner, A. Ody, F. Poulet, *Science* **2014**, *343*, 1343-1346.
[119] S. Marchi, W. F. Bottke, L. T. Elkins-Tanton, M. Bierhaus, K. Wuennemann, A. Morbidelli, D. A. Kring, *Nature* **2014**, *511*, 578-582.
[120] O. Abramov, S. J. Mojzsis, *Nature* **2009**, *459*, 419-422.
[121] T. Kleine, M. Touboul, B. Bourdon, F. Nimmo, K. Mezger, H. Palme, S. B. Jacobsen, Q. Z. Yin, A. N. Halliday, *Geochim. Cosmochim. Ac.* **2009**, *73*, 5150-5188.
[122] P. Boehnke, T. M. Harrison, *Proc. Natl. Acad. Sci. U.S.A.* **2016**, *113*, 10802-10806.

## Entry for the Table of Contents

# REVIEW

Models for Earth's impact history explain the late delivery of platinum, gold, and other siderophiles via a ~$10^{23}$ kg impactor (Moneta) ~4.48 billion years ago. The iron core from this impactor would have reduced the atmosphere above a relatively oxidized mantle, opening a window of opportunity for RNA precursor synthesis. Surprisingly, this suggests that RNA formation was most probable ~4.36 ±0.1 billion years ago.

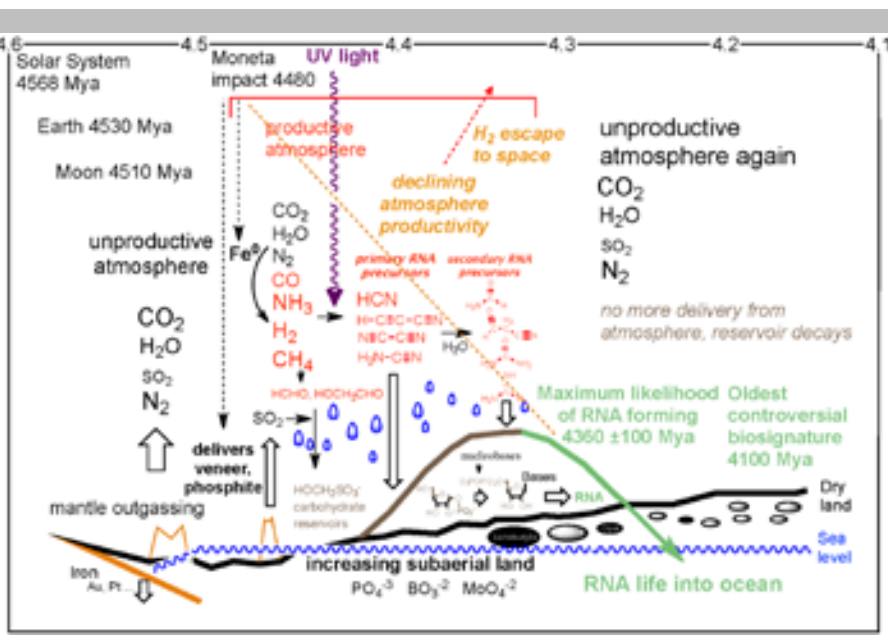

Steven A. Benner,* Elizabeth A. Bell, Elisa Biondi, Ramon Brasser, Thomas Carell, Hyo-Joong Kim, Stephen J. Mojzsis, Arthur Omran, Matthew A. Pasek, and Dustin Trail

**When did Life Likely Emerge on Earth in an RNA-First Process?**